\documentclass{aa}  

\usepackage{graphicx}
\usepackage{txfonts}
\usepackage{hyperref}
\usepackage{natbib}
\usepackage{lineno}
\begin{document} 
\bibpunct{(}{)}{;}{a}{}{,} 

\title{GammaLib and ctools}
\subtitle{A software framework for the analysis of astronomical gamma-ray data}

\author{
J. Kn\"odlseder\inst{1} \and 
M. Mayer\inst{2} \and 
C. Deil\inst{3} \and 
J.-B. Cayrou\inst{1} \and 
E. Owen\inst{3} \and 
N. Kelley-Hoskins\inst{4} \and 
C.-C. Lu\inst{3} \and 
R. Buehler\inst{4} \and 
F. Forest\inst{1} \and 
T. Louge\inst{1} \and 
H. Siejkowski\inst{5} \and 
K. Kosack\inst{6} \and 
L. Gerard\inst{4} \and 
A. Schulz\inst{4} \and 
P. Martin\inst{1} \and 
D. Sanchez\inst{7} \and 
S. Ohm\inst{4} \and 
T. Hassan\inst{8} \and 
S. Brau-Nogu\'e\inst{1}
}
\institute{
Institut de Recherche en Astrophysique et Plan\'etologie, 9 avenue Colonel-Roche,
31028 Toulouse, Cedex 4, France \and
Department of Physics, Humboldt University Berlin, Newtonstr. 15, 12489 Berlin, Germany \and
Max-Planck-Institut f\"ur Kernphysik, Saupfercheckweg 1, 69117 Heidelberg, Germany \and
Deutsches Elektronen-Synchrotron, Platanenallee 6, 15738 Zeuthen, Germany \and
AGH University of Science and Technology, ACC Cyfronet AGH, ul. Nawojki 11, PO Box 386,
PL-30-950 Krak\'{o}w 23, Poland \and
CEA/IRFU/SAp, CEA Saclay, Bat 709, Orme des Merisiers, 91191 Gif-sur-Yvette, France \and
Laboratoire d'Annecy-le-Vieux de Physique des Particules, 9 Chemin de Bellevue - BP 110, 74941 Annecy-le-Vieux Cedex, France \and
Institut de Fisica d'Altes Energies, The Barcelona Institute of Science and Technology, Campus UAB, 08193 Bellaterra, Spain
}

\date{Received April 29, 2016; Accepted June 1, 2016}

\abstract
{
The field of gamma-ray astronomy has seen important progress during the last decade, yet to 
date no common software framework has been developed for the scientific analysis of
gamma-ray telescope data. 
We propose to fill this gap by means of the GammaLib software, a generic library that we have
developed to support the analysis of gamma-ray event data.
GammaLib was written in C++ and all functionality is available in Python through an extension
module.
Based on this framework we have developed the ctools software package, a suite of software
tools that enables flexible workflows to be built for the analysis of Imaging Air Cherenkov Telescope
event data.
The ctools are inspired by science analysis software available for existing high-energy 
astronomy instruments, and they follow the modular ftools model developed by the High 
Energy Astrophysics Science Archive Research Center.
The ctools were written in Python and C++, and can be either used from the command
line via shell scripts or directly from Python. 
In this paper we present the GammaLib and ctools software versions 1.0 that were
released at the end of 2015.
GammaLib and ctools are ready for the science analysis of Imaging Air Cherenkov Telescope
event data, and also support the analysis of Fermi-LAT data and the exploitation of the COMPTEL
legacy data archive.
We propose using ctools as the Science Tools software for the Cherenkov Telescope Array
Observatory.
}

\keywords{
methods: data analysis --
virtual observatory tools
}

\maketitle

\section{Introduction}
\label{sec:intro}

The last decade has seen important progress in the field of gamma-ray astronomy thanks to
significant improvements in the performance of ground-based and space-based gamma-ray 
telescopes.
Gamma-ray photons are currently studied over more than eight decades in energy, from a few
100 keV up to more than 10 TeV.
The technologies used for observing gamma rays are very diverse and cover indirect imaging
devices, such as coded mask and Compton telescopes, and direct imaging devices, such as
pair creation telescopes, and air or water Cherenkov telescopes.

Despite this technical diversity, the high-level data that is produced by the instruments for scientific
analysis have great similarities.
Generally, the data is comprised of lists of individual events, each of which is characterised by
temporal, directional and energy information.
Many space-based missions (e.g. CGRO, INTEGRAL, Fermi) provide event data using the Flexible
Image Transport System (FITS) data format \citep{pence2010} and follow the Office of Guest
Investigator Programs (OGIP) conventions \citep{corcoran1995}, and efforts are also underway
to convert data from existing ground-based observatories (H.E.S.S., VERITAS, MAGIC, 
HAWC) into the same framework.\footnote{\url{http://gamma-astro-data-formats.readthedocs.org}}
The next-generation ground-based Cherenkov Telescope Array (CTA) Observatory, a world-wide
project to create a large and sustainable Imaging Air Cherenkov Telescope (IACT) observatory
\citep{acharya2013}, will also follow that path.
The CTA will provide all-sky coverage by implementing two IACT arrays, one in the northern
hemisphere that will be located in La Palma, and one in the southern hemisphere that will be
located in Chile, equipped in total with more than a hundred IACTs of three different size classes
to cover the energy range 20~GeV -- 300~TeV.
The CTA Observatory will distribute the high-level event data in form of FITS files to the astronomical
community \citep{knoedlseder2015}.

Despite the standardisation in the data formats, there are no common tools yet for the 
scientific analysis of gamma-ray data.
So far, each instrument comes with its suite of software tools, which often requires costly
development cycles and maintenance efforts, and which puts the burden on the astronomer to 
learn how to use each of them for scientific research.
For X-ray astronomy, the High Energy Astrophysics Science Archive Research Center (HEASARC)
has developed standards (such as XSELECT or XSPEC) that unify the data analysis tasks,
making X-ray data more accessible to the astronomical community at large.

We aim to follow a similar path for gamma-ray astronomy, and therefore developed the
GammaLib software package as a general framework for the analysis of gamma-ray event
data \citep{knoedlseder2011}.
On top of this framework we developed the ctools software package, a suite of software tools that
enables building flexible workflows for the analysis of IACT event data \citep{knoedlseder2016}.
The development of the ctools and GammaLib software packages has recently been driven by the
needs for CTA, but the interest in these packages is also growing within the collaborations of existing
IACTs.
We therefore want to provide a reference publication for both software packages that makes the
community at large aware of their existence, but that also describes in some detail their content.

Both software packages are developed as open source codes\footnote{
The source code can be downloaded from \url{http://cta.irap.omp.eu/ctools/download.html}
which also provides access to Mac OS X binary packages.
Up to date user documentation for both packages can be found at
\url{http://cta.irap.omp.eu/gammalib} and \url{http://cta.irap.omp.eu/ctools}.
}
under the GNU General Public License version 3.
GammaLib and ctools are version controlled using the Git system that is managed through a GitLab
front end.\footnote{\url{https://cta-gitlab.irap.omp.eu}}
The code development is managed using Redmine\footnote{\url{https://cta-redmine.irap.omp.eu}}
and we use a Jenkins-based multi-platform continued integration system to assure the integrity and
the functionality of the software.\footnote{\url{https://cta-jenkins.irap.omp.eu}}
Code quality is controlled using SonarQube.\footnote{\url{https://cta-sonar.irap.omp.eu}}
GammaLib and ctools are built and tested on a large variety of Linux distributions, on Mac OS X
(10.6 -- 10.11), on FreeBSD and on OpenSolaris.
Usage on Windows is not supported.

GammaLib is mostly written in C++ and provides Application Programming Interfaces (APIs) for
C++ and Python.
GammaLib is an object oriented library that comprises over 200 classes for a total volume of nearly
120\,000 lines of code.
The ctools package is written in Python and C++ and comprises nearly 30 analysis tools that make
up almost 20\,000 lines of code.
The interface between C++ code and Python is generated for both packages using the Simplified
Wrapper and Interface Generator (SWIG).\footnote{\url{http://www.swig.org}}
Both Python 2 (from version 2.3 on) and Python 3 are supported.

This paper describes GammaLib and ctools release version 1.0.
Section \ref{sec:gammalib} and section \ref{sec:ctools} present the GammaLib and ctools software
packages, respectively.
Section \ref{sec:performance} summarises the performance of the software.
We conclude with an outlook on future developments in Sect. \ref{sec:outlook}.

\section{GammaLib}
\label{sec:gammalib}

\subsection{Overview}

GammaLib is a single shared library that contains all C++ classes, support functions, and
some global variables that are needed to analyse gamma-ray event data.
A central feature of GammaLib is that all functionalities that are necessary for the analysis
of gamma-ray event data are implemented natively, reducing thus external dependencies to a strict
minimum.
This is essential to keep the long-term software maintenance cost down, and helps in assuring
independence from operating systems and user-friendly installation.
The price to pay for this feature was a larger initial development effort that we had to invest at the
beginning of the project.
The only external library GammaLib relies on is HEASARC's {\em cfitsio} library\footnote{ 
\url{http://heasarc.gsfc.nasa.gov/fitsio/}}
that is however available as binary package on all modern Linux and Mac OS X systems.
An optional dependency is the {\em readline} and {\em ncurses} libraries that enhance the 
Image Reduction and Analysis Facility (IRAF) command line parameter interface that is
implemented as a user interface, but GammaLib is also fully functional if these libraries
are not available.

All GammaLib classes start with an capital ``{\tt G}", followed by a capitalised class name.
If the class name is composed of several words, each word is capitalised (CamelCase).
Examples of GammaLib class names are {\tt GModels}, {\tt GSkyMap} or
{\tt GFitsBinTable}.
Names of classes that contain lists of objects (container classes) are generally formed by
appending an ``{\tt s}" to the class name of the objects they contain.
For example, {\tt GModels} is a container class for {\tt GModel} objects.
GammaLib functions or global variables are defined within the {\tt gammalib} namespace.
For example, {\tt gammalib::expand\_env()} is a function that expands the environment
variables in a string, or {\tt gammalib::MeV2erg} is a multiplier that converts energies from
units of MeV to units of ergs.

\subsection{Software layout}

\begin{figure*}
\centering
\includegraphics[width=12cm]{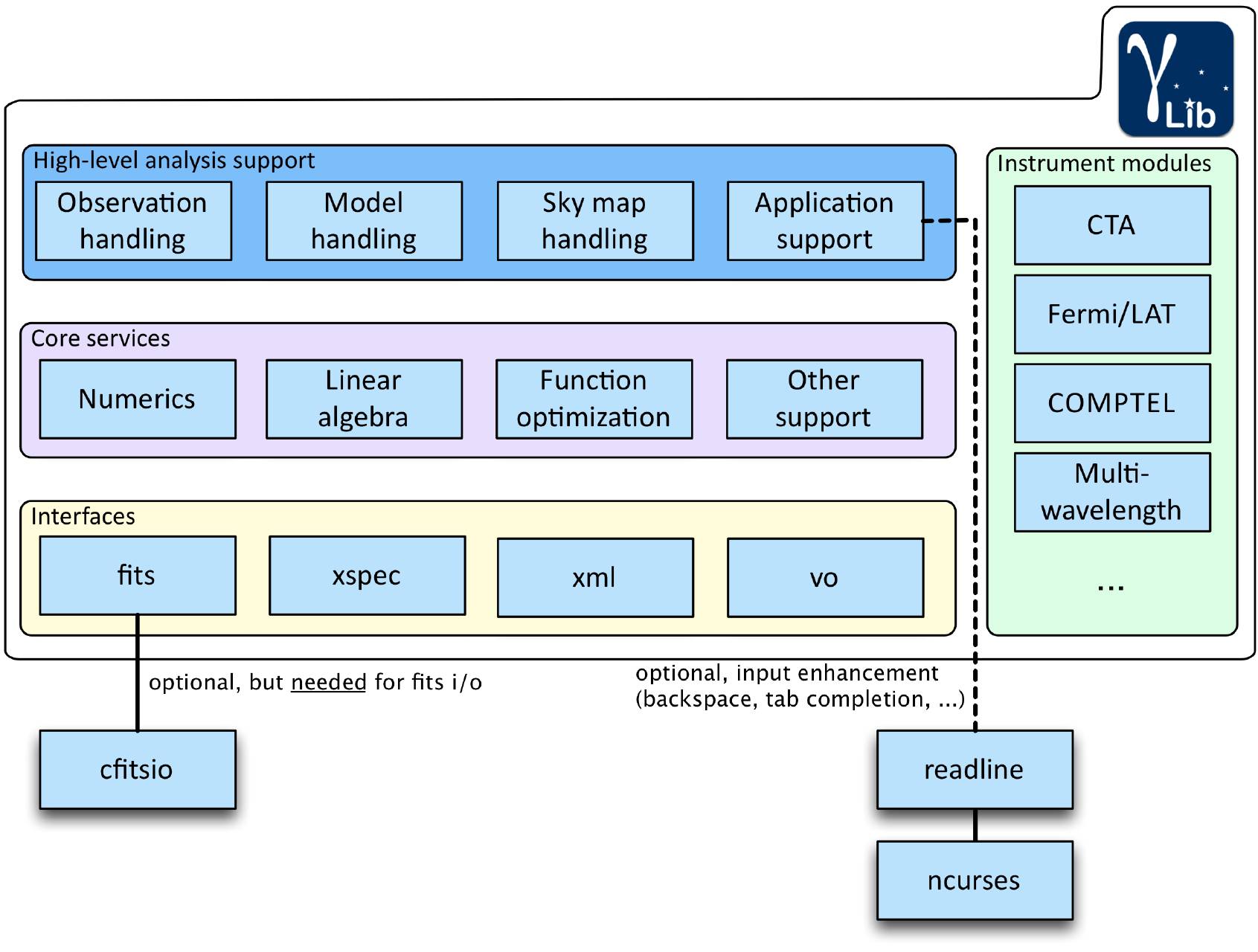}
\caption{
Organisation scheme of the GammaLib library (see text for a description of the entities shown).
\label{fig:gammalib}
}
\end{figure*}

The GammaLib classes are organised into three software layers, each of which comprises several
modules (see Fig.~\ref{fig:gammalib}).
The top layer provides support for instrument-independent high-level data analysis, comprising the
handling of observations, models, and sky maps.
Also the support for creating ftool applications is part of this layer.
Core services related to numerical computations and function optimisation are implemented in the
second layer.
The third layer is an interface layer that allows handling of data in FITS, XSPEC and
XML formats, and that implements support for Virtual Observatory (VO) interoperability.

\subsubsection{Observation module}
\label{sec:obs}

The {\em observation} module contains the abstract {\tt GObservation} base class that defines the
interface for a generic gamma-ray observation.
In GammaLib, an observation is defined as a period in time during which an instrument was taking
data, in a stable configuration that can be described by a fixed instrument response function (the time
period does not need to be contiguous).
The data is represented by events that may either be provided in a list or that are binned in a
n-dimensional data cube.

Each event or event bin is characterised by three fundamental properties:
instrument direction $\vec{p}'$,
measured energy $E'$, and
trigger time $t'$
(we use in the following primed symbols to denote reconstructed or measured quantities, and
unprimed symbols to denote true quantities).
We note that the instrument direction is not necessarily given in sky coordinates, but could be
for instance the detector number or any other instrument related property that characterises the
arrival direction of an event.
While energy and time are handled in a unit or reference independent way, it should be noted
that GammaLib stores energies internally in MeV and defines the time zero at 
1st January 2010, 00:00:00 (TT).

The {\em observation} module contains also the abstract {\tt GResponse} base class that
represents the instrument response function
$R(\vec{p}', E', t' | \vec{p}, E, t)$
that describes the transformation from the physical properties of a photon
(sky direction $\vec{p}$, energy $E$, and time $t$) to the measured characteristics of an event
(the instrument response function is given in units of cm$^2$ sr$^{-1}$ s$^{-1}$ MeV$^{-1}$).
The instrument response function and the events are the basic constitutents of a GammaLib
observation, which implies that every observation can hold its proper (and independent) instrument
response function.
We will use in the following the index $i$ to indicate that a function applies to a specific
observation.

A fundamental function for each observation is the likelihood function $L_i(M)$ that quantifies
the probability that the data collected during a given observation is drawn from a particular
model $M$ (see section \ref{sec:model} for a description of the model handling in GammaLib).
The formulae used for the likelihood computation depend on the type of the data (binned or
unbinned) and the assumed underlying statistical law.
For event lists, the Poisson formula
\begin{equation}
-\ln L_i(M) = e_i(M) - \sum_k \ln P_{i}(\vec{p}'_k, E'_k ,t'_k | M)
\label{eq:unbinned}
\end{equation}
is used, where the sum is taken over all events $k$, characterised by the instrument direction
$\vec{p}'_k$, the measured energy $E'_k$ and the trigger time $t'_k$.
$P_{i}(\vec{p}',E',t' | M)$ is the probability density that given the model $M$, an event with
instrument direction $\vec{p}'$, measured energy $E'$ and trigger time $t'$ occurs.
$e_i(M)$ is the total number of events that are predicted to occur during an
observation given the model $M$, computed by integrating the probability density over
the trigger time, measured energy and instrument direction:
\begin{equation}
e_i(M) = \int_{GTI} \int_{Ebounds} \int_{ROI} P_{i}(\vec{p}',E',t' | M) \,
\mathrm{d}\vec{p}' \, \mathrm{d}E' \, \mathrm{d}t'.
\end{equation}
The temporal integration boundaries are defined by so-called Good Time Intervals (GTIs)
that define contiguous periods in time during which data was taken.
The spatial integration boundaries are defined by a so-called Region of Interest (ROI).

For binned data following a Poisson distribution the formula
\begin{equation}
- \ln L_i(M) = \sum_k e_{k,i}(M) - n_{k,i} \ln e_{k,i}(M)
\label{eq:binned}
\end{equation}
is used, where the sum over $k$ is now taken over all data cube bins.
$n_{k,i}$ is the number of events in bin $k$ observed during observation $i$, and
\begin{equation}
e_{k,i}(M) = P_{i}(\vec{p}'_k, E'_k ,t'_k | M) \times \Omega_k \times \Delta E_k \times \Delta T_k
\end{equation}
is the predicted number of events from model $M$ in bin $k$ of observation $i$.
The probability density is evaluated for the reference instrument direction $\vec{p}'_k$,
measured energy $E'_k$ and trigger time $t'_k$ of bin $k$, typically taken to be the values at the
bin centre, and multiplied by the solid angle $\Omega_k$, the energy width $\Delta E_k$ and the
exposure time (or ontime) $\Delta T_k$ of bin $k$.\footnote{
Any dead time correction is taken into account by the instrument response function.}
Alternatively, if the data follow a Gaussian distribution the formula
\begin{equation}
- \ln L_i(M) = \frac{1}{2} \sum_k \left( \frac{n_{k,i} - e_{k,i}(M)}{\sigma_{k,i}} \right)^2
\label{eq:gaussian}
\end{equation}
is used, where $\sigma_{k,i}$ is the statistical uncertainty in the measured number of events
for bin $k$.

Observations are collected in the {\tt GObservations} container class, which is the central object
that is manipulated in a GammaLib data analysis.
By summing for a given model $M$ over the negative log-likelihood values of all observations in
the container using
\begin{equation}
- \ln L(M) = - \sum_i \ln L_i(M),
\label{eq:jointlogL}
\end{equation}
the joint maximum likelihood is computed, enabling the combination of an arbitrary number of
observations to constrain the parameters of a model $M$.
This opens the possibility of performing multi-instrument and multi-wavelength analyses of event 
data with GammaLib by combining observations performed by different instruments in a single
observation container.
If OpenMP support is available, Eq.~(\ref{eq:jointlogL}) will be parallelised
and hence benefits from the availability of multi-core and/or multi-processor infrastructures
to speed up computations.

\subsubsection{Model module}
\label{sec:model}

The {\em model} module collects all classes needed to describe event data in a parametrised way.
The generic abstract base class of all models is the {\tt GModel} class, which provides methods
to compute the probability density $P_{i}(\vec{p}',E',t' | M_j)$ of observing an event with instrument
direction $\vec{p}'$, measured energy $E'$, and trigger time $t'$ during an observation $i$ for a
given model $M_j$.
We note that we used here the index $j$ to indicate a specific model.
Models can be combined using the {\tt GModels} container class, which in turn computes the sum
\begin{equation}
P_{i}(\vec{p}',E',t' | M) = \sum_j P_{i}(\vec{p}',E',t' | M_j)
\end{equation}
of the probability densities of all models.
{\tt GModels} is a member of the {\tt GObservations} container class, so that all information
needed for gamma-ray event data analysis is contained in a single object.

There are two basic classes that derive from the abstract {\tt GModel} base class:
the {\tt GModelSky} class which implements a factorised representation of the spatial, spectral, and
temporal components of a celestial source, and the abstract {\tt GModelData} base class which
defines the interface for any instrument specific description of events, and which is generally used 
to model instrumental backgrounds.
While a celestial model $M^S_j(\vec{p}, E, t)$ is defined as function of true quantities, a data model
$M^D_{j,i}(\vec{p'}, E', t')$ is defined as function of reconstructed or measured quantities, and may
also depend on the observation $i$.
The event probability density for a celestial source model is computed by convolving the model with
the instrument response function of the observation using
\begin{equation}
P_{i}(\vec{p}', E', t' | M_j) = 
\int_{\vec{p}, E, t} R_i(\vec{p}', E', t' | \vec{p}, E, t) \times M^S_j(\vec{p}, E, t) 
\, \mathrm{d}\vec{p} \, \mathrm{d}E \, \mathrm{d}t
\end{equation}
while for a data model, the event probability density is directly given by the model
using
\begin{equation}
P_{i}(\vec{p}', E', t' | M_j) = M^D_{j,i}(\vec{p'}, E', t').
\end{equation}

The factorisation of a celestial source model is given by
\begin{equation}
M^S(\vec{p}, E, t) = M_\mathrm{S}(\vec{p} | E, t) \times M_\mathrm{E}(E | t) \times M_\mathrm{T}(t),
\label{eq:celmodel}
\end{equation}
where $M_\mathrm{S}(\vec{p} | E, t)$, $M_\mathrm{E}(E | t)$, and $M_\mathrm{T}(t)$ are the spatial,
spectral, and temporal components of the model (we drop the indices $j$ from now on).
We note that this definition allows for energy- and time-dependent spatial model components,
and for time-dependent spectral model components.
So far, however, only model components that are constant in time are implemented.

The spatial component can be either modelled as a point source, a radially symmetric source, 
an elliptical source, or a diffuse source.
For the latter, options comprise
an isotropic intensity distribution on the sky,
an arbitrary intensity distribution in form of a sky map, or
an arbitrary energy-dependent intensity distribution provided in form of a map cube.
Several model components exist for radial or elliptical sources.
Disk models describe uniform intensity distributions within radial or elliptical boundaries.
The radial Gaussian model represents an intensity distribution given by
\begin{equation}
M_\mathrm{S}(\vec{p} | E, t) = \frac{1}{2 \pi \sigma^2} \exp \left(-\frac{1}{2}\frac{\theta^2}{\sigma^2} \right),
\end{equation}
where
$\theta$ is the angular separation from the centre of the distribution, and
$\sigma$ is the Gaussian width of the distribution.
The radial shell model represents a spherical shell projected on the sky given by
\begin{equation}
M_\mathrm{S}(\vec{p} | E, t) = n_0 \left \{
 \begin{array}{l}
      \displaystyle
      \sqrt{ \sin^2 \theta_\mathrm{out} - \sin^2 \theta } -
      \sqrt{ \sin^2 \theta_\mathrm{in}  - \sin^2 \theta } \\
      \makebox[6.3cm][r]{if $\theta \le \theta_{\rm in}$} \\
      \\
     \displaystyle
       \sqrt{ \sin^2 \theta_\mathrm{out} - \sin^2 \theta } \\
      \makebox[6.3cm][r]{if $\theta_\mathrm{in} < \theta \le \theta_\mathrm{out}$} \\
      \\
     \displaystyle
     0 \\
     \makebox[6.3cm][r]{if $\theta > \theta_\mathrm{out}$}
   \end{array}
\right .
\end{equation}
where $\theta_\mathrm{in}$ and $\theta_\mathrm{out}$ are the apparent inner and outer shell radii 
on the sky, respectively, and
\begin{multline}
n_0 = \frac{1}{2 \pi} \left(
     \frac{\sqrt{1-\cos 2 \theta_\mathrm{out}} -
           \sqrt{1-\cos 2 \theta_\mathrm{in}}}{2 \sqrt{2}} \right. \\
     \left. + \frac{1+\cos 2 \theta_\mathrm{out}}{4} 
     \ln \left(
           \frac{\sqrt{2} \cos \theta_\mathrm{out}}
                {\sqrt{2} + \sqrt{1 - \cos 2 \theta_\mathrm{out}}} \right) \right. \\
    \left. - \frac{1+\cos 2 \theta_\mathrm{in}}{4}
    \ln \left(
           \frac{\sqrt{2} \cos \theta_\mathrm{in}}
                {\sqrt{2} + \sqrt{1 - \cos 2 \theta_\mathrm{in}}} \right) \right)^{-1}
\end{multline}
is a normalisation constant.
Finally, the elliptical Gaussian model represents an intensity distribution given by
\begin{equation}
M_\mathrm{S}(\theta, \phi | E, t) = n_0 \times \exp \left( -\frac{\theta^2}{2 r_\mathrm{eff}^2} \right),
\end{equation}
where the effective ellipse radius $r_\mathrm{eff}$ towards a given position angle is given by
\begin{equation}
r_\mathrm{eff} = \frac{ab} {\sqrt{\left( a \sin (\phi - \phi_0) \right)^2 + \sqrt{\left( b \cos (\phi - \phi_0) \right)^2}}}
\end{equation}
and
$a$ is the semi-major axis of the ellipse,
$b$ is the semi-minor axis,
$\phi_0$ is the position angle of the ellipse, counted counterclockwise from North, and
$\phi$ is the azimuth angle with respect to celestial North.
The normalisation constant $n_0$ is given by
\begin{equation}
n_0 = \frac{1}{2 \pi \times a \times b}.
\end{equation}

The spectral model components $M_\mathrm{E}(E | t)$ include a power law model
\begin{equation}
M_\mathrm{E}(E | t) = k_0 \left( \frac{E}{E_0} \right)^{\gamma}
\end{equation}
where $k_0$ is a prefactor, $E_0$ is the pivot energy, and $\gamma$ is the
spectral index.
We note that in GammaLib spectral indices are defined including the sign, hence typical gamma-ray
sources have values of $\gamma=-2 \ldots -4$.
A variant of the power law model that replaces the pivot energy and prefactor by the integral flux
$N$  over an energy range $[E_\mathrm{min}, E_\mathrm{max}]$ is given by
\begin{equation}
M_\mathrm{E}(E | t) = \frac{N(\gamma+1)E^{\gamma}}
                         {E_\mathrm{max}^{\gamma+1} - E_\mathrm{min}^{\gamma+1}}.
\end{equation}
A broken power law is implemented by
\begin{equation}
M_\mathrm{E}(E | t) = k_0 \times \left \{
    \begin{array}{l l}
      \left( \frac{E}{E_\mathrm{b}} \right)^{\gamma_1} & \mathrm{if\,\,} E < E_\mathrm{b} \\
      \left( \frac{E}{E_\mathrm{b}} \right)^{\gamma_2} & \mathrm{otherwise}
    \end{array}
    \right .
\end{equation}
with $E_\mathrm{b}$ being the break energy, and $\gamma_1$ and $\gamma_2$ being
the spectral indices before and after the break, respectively.
An exponentially cut-off power law is implemented by
\begin{equation}
M_\mathrm{E}(E | t) = k_0 \left( \frac{E}{E_0} \right)^{\gamma}
                    \exp \left( \frac{-E}{E_{\rm cut}} \right)
\end{equation}
with $E_\mathrm{cut}$ being the cut-off energy.
A variant of this model is the super exponentially cut-off power law model, defined by
\begin{equation}
M_\mathrm{E}(E | t) = k_0 \left( \frac{E}{E_0} \right)^{\gamma}
                    \exp \left( -\left( \frac{E}{E_\mathrm{cut}} \right)^{\alpha}
                    \right)
\end{equation}
which includes an additional power law index $\alpha$ on the cut-off term.
A log parabola model is defined by
\begin{equation}
M_\mathrm{E}(E | t) = k_0 \left( \frac{E}{E_0} \right)^{\gamma+\eta \ln(E/E_0)},
\end{equation}
where $\eta$ is a curvature parameter.
And a Gaussian function that can be used to model gamma-ray lines is defined by
\begin{equation}
M_\mathrm{E}(E | t) = \frac{N_0}{\sqrt{2\pi}\sigma}
                    \exp \left( \frac{-(E-\bar{E})^2}{2 \sigma^2} \right),
\end{equation}
where $\bar{E}$ is the centre energy, and $\sigma$ is the line width.
An arbitrary spectral model is defined by a file function based on energy and intensity values
specified in an ASCII file from which a piece-wise power law model is constructed.
The file function can be adjusted to the data by applying a global scaling factor.
Alternatively, the spectral node model is allowing for adjusting each pair of energy and intensity
values as free parameters of a piece-wise power law model.

\subsubsection{Sky map module}

The {\em sky map} module collects classes that are used for handling sky maps.
The central class of the module is the {\tt GSkyMap} class which transparently handles sky maps
provided either in the FITS World Coordinate Systems \citep[WCS;][]{greisen2002} or all-sky maps 
that are defined on the HEALPix grid \citep{gorski2005}.
Currently, seven WCS projections are implemented (Aitoff, zenithal/azimuthal perspective,
cartesian, Mercator's, Mollweide, stereographic and gnomonic), and we plan to implement in
the future the full set of projections that is available in the {\em wcslib} library.\footnote{
\url{http://www.atnf.csiro.au/people/mcalabre/WCS/wcslib/}}
We note that sky pixels in GammaLib are defined at the bin centre, hence the first pixel of a WCS map
covers the pixel range $[-0.5,+0.5]$ in the x- and the y-direction.

The {\em sky map} module also contains classes to define and handle arbitrary regions on the sky.
Methods exist to test whether a given sky direction is contained in a sky region, or whether a
region overlaps or is contained in another sky region.
The format for defining sky regions is compatible with that used by DS9,\footnote{
\url{http://ds9.si.edu/doc/ref/region.html}} enabling the loading of DS9 region files into
GammaLib.
So far, only a circular sky region has been implemented.

\subsubsection{Application module}
\label{sec:application}

The last module of the high-level analysis support layer is the {\em application} module that provides
classes that support building of ftools-like analysis tools, and which specifically establish the
link to the ctools software package (see section \ref{sec:ctools}).
The central class of this module is the {\tt GApplication} base class, which upon construction
grants access to user parameters provided in the IRAF command line parameter interface,
a format that is already widely used for high-energy astronomy analysis frameworks, including
ftools, the Chandra CIAO package, the INTEGRAL OSA software, or the Fermi-LAT Science
Tools (see Appendix \ref{sec:iraf}).

The {\tt GApplication} class also contains a logger that assures that all information provided during
the execution of a tool will be presented to the user in a uniform way.
By default, the logger will write output into an ASCII file, but simultaneous logging into the console
can be enabled upon request.

\subsubsection{Core modules}
\label{sec:core}

A number of services that are central to many GammaLib classes are collected into four core
modules (cf.~Fig.~\ref{fig:gammalib}).
Services useful for numerical computations are in the {\em numerics} module, and include
classes for numerical integration and differentiation, as well as mathematical functions
(e.g. error function, gamma function) and constants (e.g. $\pi$, $\ln 2$, $\sqrt{2}$).
Classes that allow vector and matrix operations are collected in the {\em linear algebra} module,
including classes to manage symmetric or sparse matrices.
Classes that are used for function minimisation are collected in the {\em optimisation} module.
Function minimisation is done using an optimiser, which will adjust the parameters of a function
to minimise the function value.
The standard optimiser for GammaLib is based on the iterative Levenberg-Marquardt method
\citep{marquardt1963}.
There is provision for including alternative optimisers.
Additional services are collected in the {\em support} module.
This includes classes for
linear and bilinear interpolation,
random number generation using a high-quality long-period natively implemented generator
\citep{marsaglia1994},
the handling of comma-separated value tables in ASCII files, and
filename handling.

\subsubsection{Interfaces}

GammaLib provides a number of interfaces to support input and output of files and
information.
This includes in particular an interface to FITS files which is implemented in the
{\em fits} module.
The central class of this module is {\tt GFits} which provides an in-memory representation of
a FITS file.
Each FITS file contains a list of Header Data Units (HDUs) composed of header
keywords and  either an image or a table.
Both ASCII and binary tables are supported.
To read and write FITS files, GammaLib relies on the {\em cfitsio} library.

GammaLib also includes an interface that allows manipulating of data provided in the
XSPEC format used for X-ray astronomy \citep{arnaud1996}.
GammaLib also includes a module that enables input and output of ASCII files in Extensible
Markup Language (XML) format.
Finally, GammaLib includes a module than enables the exchange of data and information
with other Virtual Observatory (VO) compatible applications, such as for example the {\em Aladin} 
interactive sky atlas that can be used for sky map display.
VO support is still experimental in the release 1.0.

\subsubsection{Instrument modules}

All modules that have been described so far are completely instrument independent, and
are used to handle data obtained with any kind of gamma-ray telescope.
To support the analysis of a specific instrument, instrument-specific modules have been
added that implement a number of pure virtual base classes of the {\em observation} module.
The general naming convention for instrument-specific classes is to prefix the class names after
the initial ``{\tt G}" with a unique instrument code, i.e. {\tt GCTAObservation} for the implementation
of the CTA observation class.
The GammaLib package includes so far instrument-specific modules for 
CTA (code {\tt CTA}, section \ref{sec:cta}),
Fermi-LAT (code {\tt LAT}, section \ref{sec:lat}), and
COMPTEL (code {\tt COM}, section \ref{sec:com}),
as well as a generic interface for multi-wavelength data (code {\tt MWL}, section \ref{sec:mwl}).
In the following sections we will describe the instrument-specific modules that exist
so far in the GammaLib package.

\subsubsection{CTA module}
\label{sec:cta}

While the initial driver for developing the CTA module was to support event data analysis for
the upcoming Cherenkov Telescope Array Observatory, the recent developments were also
motivated by the needs of existing IACTs, such as H.E.S.S., VERITAS, and MAGIC.
Maximum likelihood techniques are conventionally used for the analysis of data from
medium-energy and high-energy instruments \citep[e.g.][]{mattox1996,deboer1992, diehl2003},
but the technique is rather new in the field of very-high-energy astronomy, and consequently its
validation by applying it to data obtained with existing IACTs is extremely valuable for the
preparation of CTA.
To enable the joint analysis of data from all IACTs, specific instrument codes have been
implemented ({\tt CTA}, {\tt HESS}, {\tt VERITAS}, and {\tt MAGIC}) and should be used in the
observation definition XML file (see Appendix \ref{sec:obsxml}), although all IACTs will make 
use of the same classes of the CTA module.

CTA event data is provided as a FITS file, containing for each event
the reconstructed photon arrival direction,
the reconstructed energy,
and the arrival time.
In addition, each event is enumerated by an identifier, and optionally may be characterised
by instrument coordinates and an event phase (in case the data has been folded
with the ephemerides of a pulsar or a binary system).

The instrument response for CTA is assumed to factorise into
\begin{multline}
R(\vec{p}', E', t' | \vec{p}, E, t) = A_\mathrm{eff}(\vec{p}, E, t) \times
\mathrm{\it PSF}(\vec{p}' | \vec{p}, E, t) \\
\times E_\mathrm{disp}(E' | \vec{p}, E, t)
\label{eq:ctarsp}
\end{multline}
where
$A_\mathrm{eff}(\vec{p}, E, t)$ is the effective area in units of cm$^2$,
$\mathrm{\it PSF}(\vec{p}' | \vec{p}, E, t)$ is the point spread function that satisfies
\begin{equation}
\int \mathrm{\it PSF}(\vec{p}' | \vec{p}, E, t) \, \mathrm{d}\vec{p}' =1
\end{equation}
and $E_\mathrm{disp}(E' | \vec{p}, E, t)$ is the energy dispersion that satisfies
\begin{equation}
\int E_\mathrm{disp}(E' | \vec{p}, E, t) \, \mathrm{d}E'  = 1.
\end{equation}
In addition, the expected instrumental background rate
$B_\mathrm{rate}(\vec{p}', E', t')$, given in units of counts s$^{-1}$~MeV$^{-1}$~sr$^{-1}$,
is treated as the fourth response component.

\begin{figure}
\centering
\includegraphics[width=9cm]{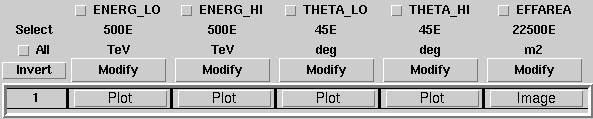}
\caption{
Screenshot of effective area information stored in {\em Response Table} format.
\label{fig:ctarsptable}
}
\end{figure}

The format for storing the CTA response information has been inspired by the FITS file 
format that is used by the Fermi-LAT Science Tools, and is based on so-called
{\em Response Tables}.
A {\em Response Table} consists of a FITS binary table with a single row and a number
of vector columns that stores an arbitrary number of n-dimensional data cubes of identical
size.
The axes of each cube dimension are defined using pairs of columns that specify the lower
and upper bin boundaries of all axis bins.
The names of the columns are composed by appending the suffixes {\tt \_LO} and {\tt \_HI}
to the axes names.
The axes columns are followed by an arbitrary number of data cubes.
Figure \ref{fig:ctarsptable} illustrates how an energy and off-axis dependent effective area
response is stored in this format.
The first four columns define the lower and upper bin boundaries of the
energy axis ({\tt ENERG\_LO} and {\tt ENERG\_HI}) and the
off-axis angle axis ({\tt THETA\_LO} and {\tt THETA\_HI}), followed by a
fifth column that holds the effective area data cube ({\tt EFFAREA}).

For the point spread function, two variants exist that both depend on energy and off-axis
angle, but differ in the functional form that is used to describe the PSF.
The first variant implements a superposition of three 2D Gaussian functions that are each
characterised by a width and a relative amplitude.
Alternatively, a King profile defined by
\begin{equation}
\mathrm{\it PSF}(\vec{p}' | \vec{p}, E, t) =
\frac{1}{2 \pi \sigma^2} \left( 1 - \frac{1}{\gamma} \right)
\left( 1 + \frac{1}{2 \gamma} \frac{\delta^2}{\sigma^2} \right)^{-\gamma},
\label{eq:psfking}
\end{equation}
can be used, where
$\delta$ is the angular separation between the true and measured photon directions
$\vec{p}$ and $\vec{p}'$, respectively,
$\sigma$ describes the width and
$\gamma$ the tail of the distribution.
In both cases, the energy and off-axis angle dependent functional parameters are stored in
the data cubes of the respective {\em Response Tables}.

The energy dispersion is stored as a three-dimensional data cube spanned by true energy, 
the ratio of reconstructed over true energy, and off-axis angle.
The background rate is stored as a three-dimensional data cube spanned by the 
detector coordinates {\tt DETX} and {\tt DETY} and the reconstructed energy.

In addition to the fundamental factorisation (Eq.~\ref{eq:ctarsp}) of the CTA instrument response,
there exists a specific response definition that is used in a so-called {\em stacked analysis}.
In a {\em stacked analysis}, data from multiple observations is combined into a single counts cube,
and consequently, an average response needs to be computed based on a proper weighting
of the individual instrument response functions for each observation.
The response for a stacked analysis is composed of an exposure cube, a point spread function
cube and a background cube; the handling of energy dispersion is not supported so far.
The exposure cube is computed using
\begin{equation}
X_\mathrm{cube}(\vec{p}, E) = \sum_i A_{\mathrm{eff},i}(\vec{p}, E, t) \times \tau_i,
\label{fig:ctaexpcube}
\end{equation}
where $A_{\mathrm{eff},i}(\vec{p}, E, t)$ is the effective area and
$\tau_i$ the livetime of observation $i$, and the sum is taken over all observations.
The point spread function cube is computed using
\begin{equation}
\mathrm{\it PSF}_\mathrm{cube}(\vec{p}, E, \delta) =
\frac{\sum_i \mathrm{\it PSF}_i(\vec{p}' | \vec{p}, E, t) \times A_{\mathrm{eff},i}(\vec{p}, E, t) \times \tau_i}
{\sum_i A_{\mathrm{eff},i}(\vec{p}, E, t) \times \tau_i},
\label{fig:ctapsfcube}
\end{equation}
%
%
and the background cube is computed using
\begin{equation}
B_\mathrm{cube}(\vec{p'}, E') =
\frac{\sum_i B_i(\vec{p}', E', t') \times \tau_i}{\sum_i \tau_i}.
\label{fig:ctabkgcube}
\end{equation}

The CTA module also contains some models to describe the distribution of the instrumental
background in the data.
All models are factorised using
\begin{equation}
B(\vec{p}', E', t') = B_\mathrm{S}(\vec{p}' | E', t') \times B_\mathrm{E}(E' | t') \times B_\mathrm{T}(t'),
\label{eq:ctamodels}
\end{equation}
where
$B_\mathrm{S}(\vec{p}' | E', t')$ is the spatial,
$B_\mathrm{E}(E' | t')$ the spectral, and
$B_\mathrm{T}(t')$ the temporal component of the model.
We note that this factorisation is similar to the one used for celestial source models (Eq.~\ref{eq:celmodel})
with all true quantities being replaced by reconstructed quantities.
In fact, the spectral and temporal model components $M_\mathrm{E}(E' | t')$ and $M_\mathrm{T}(t')$
that are provided by the {\em model} module can be used as spectral and temporal components of
the CTA background model, and only the spatial component needs an instrument specific
implementation (this is related to the fact that the instrument direction $\vec{p}'$ is instrument
specific, while energy $E'$ and time $t'$ are generic quantities).

The available background models for CTA differ in the implementation of the spatial component
$B_\mathrm{S}(\vec{p}' | E', t')$.
A first option consists of using the energy-dependent background rate templates that are stored
in the instrument response function to describe the spatial component of the model, i.e.
\begin{equation}
B_\mathrm{S}(\vec{p}' | E', t') = B_\mathrm{rate}(\vec{p}', E', t') \, .
\end{equation}
Alternatively, the spatial distribution of the background rate can be modelled using the effective
area of the instrument, i.e.
\begin{equation}
B_\mathrm{S}(\vec{p}' | E', t') = A_\mathrm{eff}(\vec{p'}, E', t') \, ,
\end{equation}
with the true arrival direction, energy and time being replaced by the reconstructed quantities.
A specific model exists also to model the background for a {\em stacked analysis}, using the background
cube as spatial component, i.e.
\begin{equation}
B_\mathrm{S}(\vec{p}' | E', t') = B_\mathrm{cube}(\vec{p'}, E') \, .
\end{equation}
Finally, a radially symmetric parametric background model can be used that defines the off-axis
dependence of the background rate as function of the offset angle $\theta$.
Three implementations for the radial dependency exist:
a Gaussian in offset angle squared, defined by
\begin{equation}
B_\mathrm{S}(\vec{p}' | E', t') = \exp \left( - \frac{1}{2} \frac{\theta^4}{\varsigma^2} \right),
\end{equation}
where $\varsigma$ is a width parameter, a radial profile defined by
\begin{equation}
B_\mathrm{S}(\vec{p}' | E', t') = \left( 1 + \left( \frac{\theta}{c_0} \right)^{c_1} \right)^{-c_2/c_1},
\end{equation}
where
$c_0$ is the width of the radial profile,
$c_1$ is the width of the central plateau, and
$c_2$ is the size of the tail of the radial distribution, and
a polynomial function
\begin{equation}
B_\mathrm{S}(\vec{p}' | E', t') = \sum_{l=0}^m c_l \theta^l,
\end{equation}
where
$c_l$ are polynomial coefficients, and
$m$ is the degree of the polynomial.

\subsubsection{Fermi-LAT module}
\label{sec:lat}

The Fermi-LAT module provides support to include event data collected with the Large Area Telescope
(LAT) aboard NASA's Fermi satellite into a joint maximum likelihood analysis, but it relies on the
official Fermi Science Tools\footnote{\url{http://fermi.gsfc.nasa.gov/ssc/data/analysis/}} to prepare the
data in the appropriate format.
So far, the module supports only the analysis of data processed with the Pass 6 and Pass 7
event-level reconstructions, but the implementation of Pass 8 analysis support is in progress.
Also, only a binned maximum likelihood analysis has been implemented so far.
Such an analysis requires on input
a livetime cube, prepared using the {\tt gtltcube} Fermi-LAT Science Tool, and
a source map that also includes a counts map, created using the {\tt gtsrcmaps}  Fermi-LAT
Science Tool.

The Fermi-LAT response function is factorised using
\begin{equation}
R(\vec{p}', E', t' | \vec{p}, E, t) =
A_\mathrm{eff}(E, \theta) \times \mathrm{\it PSF}(\delta | E, \theta) \times E_\mathrm{disp}(E' | E, \theta)
\end{equation}
where
$A_\mathrm{eff}(E, \theta)$ is the effective area in units of cm$^2$,
$\mathrm{\it PSF}(\delta | E, \theta)$ is the point spread function that satisfies
\begin{equation}
\int \mathrm{\it PSF}(\delta | E, \theta) \, \mathrm{d}\delta =1,
\end{equation}
$E_\mathrm{disp}(E' | E, \theta)$ is the energy dispersion that satisfies
\begin{equation}
\int E_\mathrm{disp}(E' | E, \theta) \, \mathrm{d}E'  = 1,
\end{equation}
$\theta$ is the inclination angle with respect to the LAT z-axis, and
$\delta$ is the angular separation between the true and measured photon directions
$\vec{p}$ and $\vec{p}'$, respectively.
The instrument response function is independent of time.
Two functional forms are available for the point spread function which are both composed
of a superposition of two King functions:
\begin{multline}
\mathrm{\it PSF}_1(\delta | E, \theta) =
n_\mathrm{c}
\left( 1-\frac{1}{\gamma_\mathrm{c}} \right)
\left( 1 + \frac{1}{2\gamma_\mathrm{c}} \frac{\delta^2}{\sigma^2} \right)^{-\gamma_\mathrm{c}}
\\
+ n_\mathrm{t}
\left( 1-\frac{1}{\gamma_\mathrm{t}} \right)
\left( 1 + \frac{1}{2\gamma_\mathrm{t}} \frac{\delta^2}{\sigma^2} \right)^{-\gamma_\mathrm{t}},
\end{multline}
and
\begin{multline}
\mathrm{\it PSF}_3(\delta | E, \theta) =
n_\mathrm{c}
\left(
\left( 1-\frac{1}{\gamma_\mathrm{c}} \right)
\left( 1 + \frac{1}{2\gamma_\mathrm{c}} \frac{\delta^2}{s_\mathrm{c}^2} \right)^{-\gamma_\mathrm{c}} \right.
\\
\left.
+ n_\mathrm{t}
\left( 1-\frac{1}{\gamma_\mathrm{t}} \right)
\left( 1 + \frac{1}{2\gamma_\mathrm{t}} \frac{\delta^2}{s_\mathrm{t}^2} \right)^{-\gamma_\mathrm{t}}
\right).
\end{multline}
The parameters $n_\mathrm{c}$, $n_\mathrm{t}$, $s_\mathrm{c}$, $s_\mathrm{t}$, $\sigma$,
$\gamma_\mathrm{c}$ and $\gamma_\mathrm{t}$ depend on energy $E$ and off-axis angle 
$\theta$.
The energy dispersion is so far not used.

The LAT events are partitioned into exclusive event types that for Pass 6 and Pass 7
data correspond to pair conversions located in either the front or the back section of the tracker
(for Pass 8 the event partitioning has been generalised to other event types).
For each event type, a specific response function exists that will be designated in the following
with the superscript $\alpha$.

The livetime cube is a means to speed up the exposure calculations in a Fermi-LAT analysis
and contains the integrated livetime as a function of sky position and inclination angle with respect
to the LAT z-axis.
This livetime, denoted by $\tau(\vec{p}, \theta)$, is the time that the LAT observed a given position
on the sky at a given inclination angle, and includes the history of the LAT's orientation during the
entire observation period.
A Fermi-LAT livetime cube includes also a version of the livetime information that is weighted by the 
livetime fraction (i.e. the ratio between livetime and ontime) and that allows correction of inefficiencies
introduced by so-called ghost events, and that we denote here by $\tau_\mathrm{wgt}(\vec{p}, \theta)$.
The exposure for a given sky direction $\vec{p}$, photon energy $E$ and event type $\alpha$ is then
computed using
\begin{multline}
X^\alpha(\vec{p}, E) =
f_1^\alpha(E) \int_{\theta} \tau(\vec{p}, \theta) \, A_\mathrm{eff}^\alpha(E, \theta) \, \mathrm{d}\theta
\\
+ f_2^\alpha(E) \int_{\theta} \tau_\mathrm{wgt}(\vec{p}, \theta) \, A_\mathrm{eff}^\alpha(E, \theta) \, \mathrm{d}\theta ,
\end{multline}
and the exposure weighted point spread function is computed using
\begin{multline}
\mathrm{\it PSF}^\alpha(\delta | \vec{p}, E) =
f_1^\alpha(E) \int_{\theta} \tau(\vec{p}, \theta) \, A_\mathrm{eff}^\alpha(E, \theta)
\, \mathrm{\it PSF}^\alpha(\delta | E, \theta) \, \mathrm{d}\theta
\\
+ f_2^\alpha(E) \int_{\theta} \tau_\mathrm{wgt}(\vec{p}, \theta) \, A_\mathrm{eff}^\alpha(E, \theta)
\, \mathrm{\it PSF}^\alpha(\delta | E, \theta) \, \mathrm{d}\theta,
\end{multline}
where $f_1^\alpha(E)$ and $f_2^\alpha(E)$ are energy and event type dependent efficiency factors.

Finally, the point spread function for a point source is computed using
\begin{equation}
\overline{\mathrm{\it PSF}}(\delta | \vec{p}, E) = \frac
{\sum_\alpha \mathrm{\it PSF}^\alpha(\delta | \vec{p}, E)}
{\sum_\alpha X^\alpha(\vec{p}, E)},
\end{equation}
where the sum is taken over all event types.

\subsubsection{COMPTEL module}
\label{sec:com}

The COMPTEL module provides support to include event data collected with the Compton
Telescope aboard NASA's CGRO mission into a joint maximum likelihood analysis.
The module accepts high-level data available at HEASARC's archive for high-energy astronomy
missions,\footnote{\url{http://heasarc.gsfc.nasa.gov/docs/journal/cgro7.html}} and is to our knowledge
the only software that can be used today to exploit the legacy COMPTEL data in that archive.
There are three basic data files in FITS format that are used to describe a COMPTEL observation
for a given energy range and that are available from HEASARC:
a DRE file, containing the binned event data,
a DRX file, containing the exposure as function of sky direction,
and a DRG file, containing geometry information.
In addition, the energy-dependent instrument response is described by IAQ files that however
are not available in the HEASARC archive.
Therefore, IAQ files applying to the standard energy ranges of COMPTEL
($0.75-1$, $1-3$, $3-10$ and $10-30$ MeV)
are included in the GammaLib package.

COMPTEL measured photons using two detector planes separated by 1.5~m, where an incoming
photon interacts first by Compton scattering with a detector of the upper plane before being
absorbed in a detector of the lower plane.
A COMPTEL event is characterised by an instrument direction spanned by the angles
$(\chi, \psi, \bar{\varphi})$.
$(\chi, \psi)$ is the direction of the photon after scattering in the upper detector plane, which is
determined from the photon interaction locations in both detector planes, and
\begin{equation}
\bar{\varphi} = \arccos \left( 1 - \frac{m_\mathrm{e}c^2}{E_2} + \frac{m_\mathrm{e}c^2}{E_1+E_2} \right)
\end{equation}
is the Compton scattering angle as inferred from the energy deposits $E_1$ and $E_2$ in the upper
and lower detector planes, respectively.
The measured energy of the photon is estimated from the sum
\begin{equation}
E' = E_1 + E_2
\end{equation}
of the energy deposits in both detector planes.
The probability that a photon which interacted in the upper detector plane will encounter a detector
of the lower plane is described by $DRG(\chi, \psi, \bar{\varphi})$, which also includes any
cuts related to the removal of events coming from the Earth limb.

The COMPTEL response is factorised using
\begin{equation}
R(\vec{p}', E', t' | \vec{p}, E, t) =
\frac{DRX(\vec{p})}{T} \times
DRG(\chi, \psi, \bar{\varphi}) \times
IAQ(\chi, \psi, \bar{\varphi} | \vec{p}, E),
\end{equation}
where
$DRX(\vec{p})$ is the exposure in units of cm$^2$ s,
$T$ is the ontime in units of s, and
$IAQ(\chi, \psi, \bar{\varphi} | \vec{p}, E)$ quantifies the interaction probability for a Compton
scattering in the upper detector plane followed by an interaction in the lower detector plane.
We note that $IAQ(\chi, \psi, \bar{\varphi} | \vec{p}, E)$ is azimuthally symmetric about the
source direction, and the IAQ file is stored as a 2D FITS image providing the interaction
probabilities as function of $\bar{\varphi}$ and $\varphi_\mathrm{geo}$ for a given energy
range, where $\varphi_\mathrm{geo}$ is the angular separation between $(\chi, \psi)$ and
$\vec{p}$.

The {\tt GCOMObservation} class implements a COMPTEL observation for a given energy
range.
Performing a spectral analysis for COMPTEL thus implies appending an observation per energy
range to the observation container.
A COMPTEL observation also contains a DRB data cube that provides an estimate of the
instrumental background.
This instrumental background model can be adjusted to the data by maximum likelihood fitting
of its $\bar{\varphi}$ distribution.
Since DRB files are not provided by HEASARC, the user may specify the DRG file as a
first order approximation of the instrumental background distribution in COMPTEL
data.
It is planned to implement more accurate background models for COMPTEL in the future.

\subsubsection{Multi-wavelength module}
\label{sec:mwl}

The multi-wavelength module provides support to add flux points that were obtained by
external analyses to constrain a joint maximum likelihood fit.
A typical example would be to constrain a synchrotron component using data obtained
at radio wavebands or in the optical or X-ray bands.
Another example is the addition of gamma-ray flux points in case the original event data
is not publicly available.
Since the multi-wavelength module handles data in physics space, the response function
has the trivial form
\begin{equation}
R(\vec{p}', E', t' | \vec{p}, E, t) = 1.
\end{equation}
So far, flux points need to be specified in a FITS table composed of at least 2 columns.
If the table contains two columns it is assumed that they respectively contain the energy
and the flux information.
For three columns it is assumed that the third column contains the statistical uncertainty in the
flux measurement.
For four or more columns it is assumed that the first four columns respectively contain the energy,
the energy uncertainty, the flux and the flux uncertainty.
Energy (or wavelength) information can be provided in units of erg(s), keV, MeV, GeV, TeV, or
{\AA}ngstr\"om.
Flux information can be provided in units of ph~cm$^{-2}$~s$^{-1}$~MeV$^{-1}$ or
erg~cm$^{-2}$~s$^{-1}$.
The FITS table unit keywords are analysed to infer the proper units of the energy and flux
axes.
We plan to connect in the future the multi-wavelength module to the Virtual Observatory interface
to support the interoperability with VO services.

\section{ctools}
\label{sec:ctools}

\subsection{Overview}

The ctools package has been written with the goal to provide a user-friendly set of software tools
allowing for the science analysis of IACT event data.
The software operates on lists of reconstructed IACT events that have been calibrated in energy
and from which most of the particle background has been removed based on air Cherenkov
shower image characteristics.
The software also requires IACT instrument response functions describing the transformation
from physical properties of photons to measured characteristics of events.
We propose to use ctools as the Science Tools software for CTA, yet the software also supports
the analysis of data from existing IACTs such as H.E.S.S., VERITAS, or MAGIC, provided that the
data and response functions are converted into the proper format.

The ctools package allows for the creation of images, spectra and light curves of gamma-ray
sources, providing the results in FITS format that is compatible with standard astronomical
tools that can be used for their display.
Tools to graphically display the analysis results are therefore not included in ctools, but a number
of Python plotting scripts based on the {\em matplotlib} Python module \citep{hunter2007} are
available in the {\tt examples} folder of the ctools package for visualisation of results.

Each ctool performs a single, well-defined analysis step, so that scientists can combine
the modular tools into a customised workflow that matches the specific needs of an analysis.
The ctools philosophy is very similar to the rational behind the ftools \citep{pence1993}, which are
widely used in X-ray astronomy, and have also inspired the science analysis frameworks of
INTEGRAL and Fermi-LAT.

The ctools package is based on the GammaLib analysis framework, which is the only external
library dependency of the software.
This assures the seamless installation of the software on a large variety of operating systems and
platforms, and keeps the long-term software maintenance cost at a manageable level.
Each tool of the package is a class that derives from GammaLib's {\tt GApplication} class,
providing thus a standard user interface and common functionalities and behavior to all tools.
A tool can be implemented as compiled executable written in C++ or as a Python script.
To distinguish both, we call the former a {\em ctool} and the latter a {\em cscript}.
Names of ctools start with ``ct'' while names of cscripts start with ``cs''.
ctools and cscripts expose identical user interfaces, and are largely indistinguishable to the
user.
Our current philosophy is to use ctools for the basic building blocks of the package, while cscripts are
used for high-level tasks, calling eventually several of the ctools.
We will use the terms {\em tool} or {\em tools} in this paper if we make no distinction between a ctool 
or a cscript.

All tools can be called from the command line using the IRAF Command Language parameter
interface (see Appendix \ref{sec:iraf}).
All tools are also available as Python classes in the {\tt ctools} and {\tt cscripts} Python modules
that are compliant with Python 2 (from version 2.3 on) and Python 3.
Within Python, observation containers can be passed from one tool to another, avoiding the need
for storing intermediate results on disk.
This enables the creation of pure in-memory analysis workflows, circumventing potential I/O
bottlenecks and profiting from the continuously growing amount of memory that is available on
modern computers.

The ctools package ships with a calibration database that contains instrument response functions
that are needed to simulate and analyse CTA event data.
The calibration database is organised following HEASARC's calibration database (CALDB)
format,\footnote{\url{http://heasarc.gsfc.nasa.gov/docs/heasarc/caldb/caldb_intro.html}}
which places all calibration relevant information into a directory tree starting from the path defined
by the {\tt CALDB} environment variable.
Instrument response functions are specified for ctools by the {\tt caldb} and {\tt irf} parameters,
where the first gives the name of the calibration database (which is {\tt caldb=prod2} for the
IRFs shipped with ctools), and the second gives the name of the IRF
(one of {\tt North\_0.5h}, {\tt North\_5h}, {\tt North\_50h}, {\tt South\_0.5h}, {\tt South\_5h}, or
{\tt South\_50h}, labelling response functions for the northern and southern CTA arrays, with variants
that have been optimised for exposure times of 0.5 hours, 5 hours and 50 hours).

\subsection{Available tools}

\begin{figure*}
\centering
\includegraphics[width=12cm]{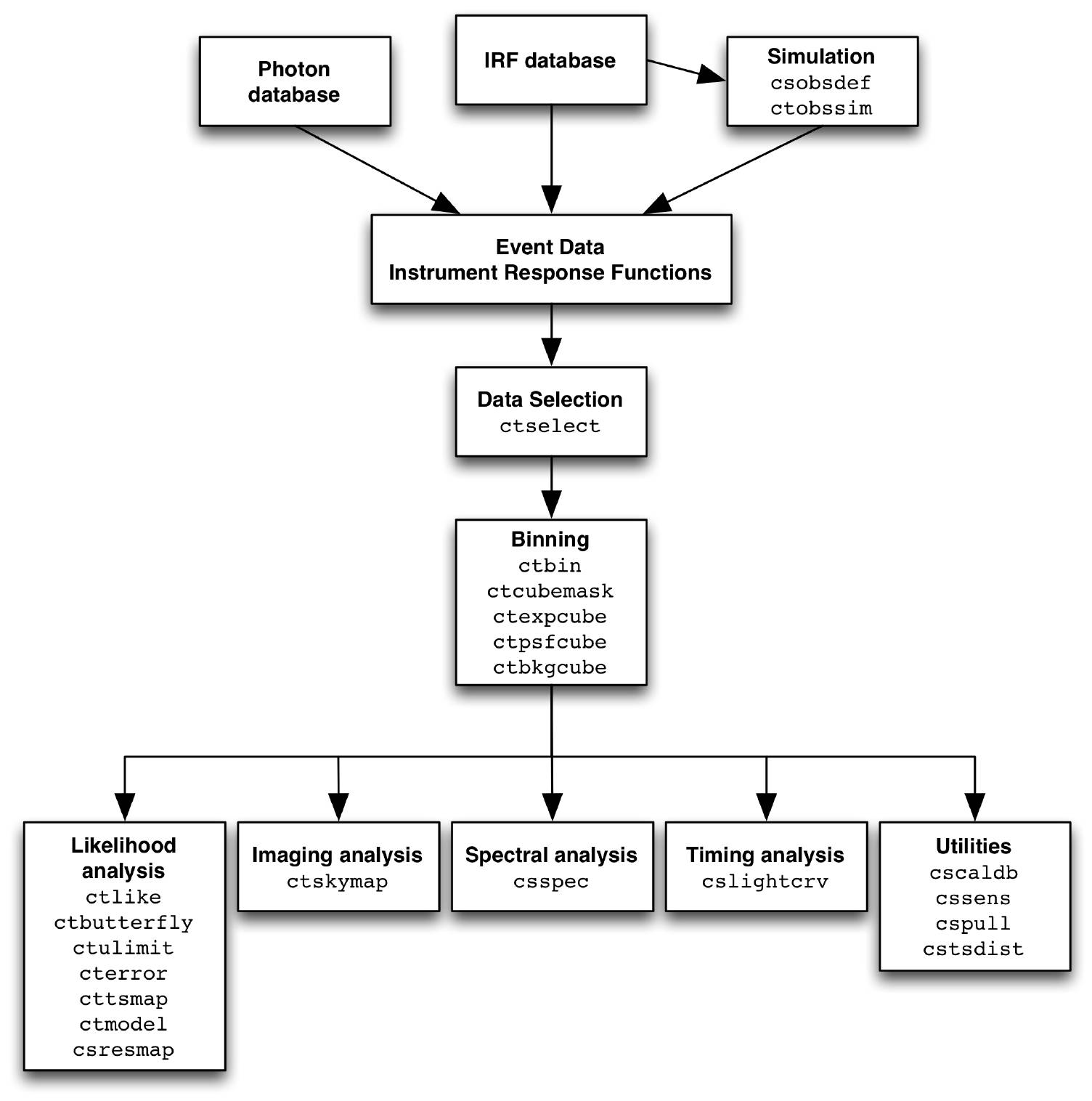}
\caption{
Overview over ctools.
\label{fig:ctools}
}
\end{figure*}

Figure \ref{fig:ctools} provides a summary of the available tools in release 1.0 of the ctools
package, grouped according to functionality, and arranged according to the typical usage in
a workflow.
Workflow examples can be found in the online ctools user manual.\footnote{
\url{http://cta.irap.omp.eu/ctools/user_manual/getting_started/quickstart.html}}
Simulation of event data is supported by the {\tt csobsdef} script and the {\tt ctobssim} tool,
event data selection is done using the {\tt ctselect} tool, and event binning and related response
preparation is supported by the {\tt ctbin}, {\tt ctcubemask}, {\tt ctexpcube}, {\tt ctpsfcube}, and
{\tt ctbkgcube} tools.
For maximum likelihood analysis there are the {\tt ctlike}, {\tt ctbutterfly}, {\tt ctulimit}, {\tt cttsmap} and
{\tt ctmodel} tools and the {\tt csresmap} script.
For imaging analysis there exists the {\tt ctskymap} tool,
spectral analysis is done using the {\tt csspec} script, and
for timing analysis there is the {\tt cslightcrv} script.
In addition, there are the utility scripts
{\tt cscaldb} to inspect the IRF database,
{\tt cssens} to determine a sensitivity curve,
{\tt cspull} to generate pull distributions, and
{\tt cstsdist} to investigate Test Statistics distributions.
In the following we provide a brief description of the tools.

\subsubsection{csobsdef}

The {\tt csobsdef} script generates an observation definition XML file from a list of pointings that
can be used as a starting point for the simulation of IACT observations (see Appendix \ref{sec:obsxml}
for a description of the XML file format).
The pointing list is a comma-separated value (CSV) ASCII file with header keywords in the first row
followed by a list of pointings, with one pointing per row.
For example, the ASCII file

{\tiny
\begin{verbatim}
name,id,ra,dec,duration,emin,emax,rad,deadc,caldb,irf
Crab,01,83.63,22.01,1800,0.1,100,5,0.95,prod2,South_0.5h
Crab,02,83.63,21.01,1800,0.2,100,5,0.95,prod2,South_0.5h
Crab,03,83.63,23.01,1800,0.3,100,5,0.95,prod2,South_0.5h
\end{verbatim}
}

\noindent will produce an observation definition XML file containing 3 observations of 1800 s
duration, wobbling around the Crab position in Declination, and with energy thresholds, as
specified by the {\tt emin} column, increasing from 0.1~TeV to 0.3~TeV.
Alternatively to Right Ascension and Declination, Galactic longitude and latitude can be
specified using the {\tt lon} and {\tt lat} keywords.
Angles are specified in units of degrees, energies in units of TeV.
Only the keywords {\tt ra} and {\tt dec} (or {\tt lon} and {\tt lat}) are mandatory.
If no {\tt duration} keyword is provided the {\tt csobsdef} script will query a value and apply that
duration to all pointings in the list.
All other keywords are optional and default values will be assumed for all observations, unless the
keyword is explicitly specified as a parameter to {\tt csobsdef}.

\subsubsection{ctobssim}

The {\tt ctobssim} tool simulates IACT event list(s) based on an input model and the instrument
characteristics described by the instrument response function(s).
The simulation includes photon events from astrophysical sources and background events
that are drawn from the input model using the numerical random number generator provided
by GammaLib.
The seed value for the random number generator can be specified through the {\tt seed}
parameter.
By default, {\tt ctobssim} simulates a single IACT pointing and produces a single event list, but
by specifying explicitly  an observation definition XML file using the {\tt inobs} parameter, the
tool can be instructed to generate an event list for each observation that is defined in the XML
file.
If OpenMP support is available, {\tt ctobssim} will parallelise the computations and
spread the simulations of multiple observations over all available computing cores.
{\tt ctobssim} creates FITS file(s) comprising the event list and their Good Time Intervals.

\subsubsection{ctselect}

The {\tt ctselect} tool selects from an event list only those events whose reconstructed arrival directions
fall within a circular acceptance region, and whose reconstructed energies and trigger times fall
within specified boundaries.
In addition, arbitrary event selection criteria can be defined by using the {\em cfitsio} row filtering
syntax.\footnote{\url{https://heasarc.gsfc.nasa.gov/docs/software/fitsio/c/c_user/node97.html}}
Optionally, {\tt ctselect} applies save energy thresholds that are specified in the effective area
component of the instrument response function via the {\tt LO\_THRES} and {\tt HI\_THRES}
FITS header keywords, or user supplied energy thresholds that are given in an observation
definition XML file through the {\tt emin} and {\tt emax} attributes (see Appendix \ref{sec:obsxml}).
If {\tt ctselect} is applied to a single event list, the tool outputs a new FITS file that only contains
the selected events.
The tool can also be applied to a list of observations by specifying an observation definition XML
file on input.
In that case, {\tt ctselect} will create one events FITS file per observation, and outputs a new
observation definition XML file that references the new event files.

\subsubsection{ctbin}

This {\tt ctbin} tool creates a counts cube that is filled with events from event list(s).
A counts cube is a three-dimensional data cube spanned by Right Ascension or Galactic longitude,
Declination or Galactic latitude, and reconstructed energy.
The events are either taken from a single event list file or from the event lists that are specified in
an observation definition XML file.
If multiple event lists are given in the observation definition XML file, the tool will loop
over all event lists and stack their events into a single counts cube.
{\tt ctbin} creates a counts cube FITS file comprising the counts cube data, the counts cube energy
boundaries, and the Good Time Intervals of all event lists that have been filled into the counts cube.

\subsubsection{ctcubemask}

The {\tt ctcubemask} tool masks out specific regions from a counts cube by setting the corresponding
bin values to $-1$, since bins with negative values will be ignored by GammaLib in a maximum
likelihood analysis.
The tool applies a spatial mask that is comprised of a circular selection region (bins outside this
region will be ignored) and a list of circular exclusion regions (bins inside these regions will be
ignored).
The circular exclusion regions are specified by an ASCII file in a format that is inspired by the
region format used by DS9.\footnote{\url{http://ds9.si.edu/doc/ref/region.html}}
Specifically, the ASCII file contains one row per exclusion region, given in the format

{\tiny
\begin{verbatim}
circle(83.63, 21.5, 0.4)
\end{verbatim}
}

\noindent where {\tt 83.63} and {\tt 21.5} are the Right Ascension and Declination of the region
centre and {\tt 0.4} is the radius (in degrees) of the exclusion circle.
The tool also only selects counts cube energy layers that are fully contained within a
specified energy interval.
{\tt ctcubemask} creates a counts cube FITS file that is a copy of the input counts cube where
all masked bins will be set to values of $-1$.
Users can of course mask additional bins of the cube manually to implement more complex
masking schemes by setting individual counts cube bins to values of $-1$.

\subsubsection{ctexpcube}

The {\tt ctexpcube} tool generates an exposure cube for a stacked maximum likelihood analysis
according to Eq.~(\ref{fig:ctaexpcube}).
An exposure cube is a three-dimensional data cube spanned by Right Ascension or Galactic longitude,
Declination or Galactic latitude, and energy, which gives the exposure as function of true sky direction
and energy.
{\tt ctexpcube} creates an exposure cube FITS file comprising the exposure cube, its energy
boundaries, and the Good Time Intervals of all observations that have been used for the computation
of the exposure cube.

\subsubsection{ctpsfcube}

The {\tt ctpsfcube} tool generates a point spread function cube for a stacked maximum likelihood
analysis according to Eq.~(\ref{fig:ctapsfcube}).
A point spread function cube is a four-dimensional data cube spanned by true Right Ascension or
Galactic longitude, true Declination or Galactic latitude, true energy, and offset angle between true
and measured arrival direction of a photon.
{\tt ctpsfcube} creates a point spread function cube FITS file comprising the point spread function
cube, its energy boundaries, and the definition of the angular separation axis.

\subsubsection{ctbkgcube}

The {\tt ctbkgcube} tool generates a background cube for a stacked maximum likelihood
analysis according to Eq.~(\ref{fig:ctabkgcube}).
A background cube is a three-dimensional data cube spanned by reconstructed Right Ascension or
Galactic longitude, Declination or Galactic latitude, and energy.
An input model is used to predict the expected number of background counts in each background
cube bin.
{\tt ctbkgcube} creates a background cube FITS file comprising the background rate per
cube bin, and the energy boundaries of the background cube.
The tool also creates an output model XML file that should be used as input for a maximum
likelihood analysis.

\subsubsection{ctlike}

The {\tt ctlike} tool is the main engine of the ctools package, allowing for the determination of the flux, 
spectral index, position or extent of gamma-ray sources using maximum likelihood model
fitting of IACT event data.
The analysis can be done using an unbinned or binned formulation of the log-likelihood function
(Eqs.~\ref{eq:unbinned}, \ref{eq:binned}, and \ref{eq:gaussian}), and the tool is able to
perform a joint maximum likelihood analysis of data collected in separate observations or
with different instruments.

Multiple observations, including data collected with different instruments, can be handled by
specifying an observation definition XML file on input (see Appendix \ref{sec:obsxml}).
Each of the observations will be kept separately and associated with its appropriate instrument
response function, as opposed to a stacked analysis where average response functions are
used.
If an observation definition XML file is provided, {\tt ctlike} will use the joint likelihood of all the
observations for parameter optimisation (see Eq.~\ref{eq:jointlogL}).

By default, {\tt ctlike} will use the Poisson statistics for likelihood computation, but for binned analysis
also Gaussian statistics can be requested.
All model parameters which are flagged by the attribute {\tt free="1"} in the model XML file, including
spatial parameters, will be adjusted by {\tt ctlike}.
For all model components $j$ for which the attribute {\tt tscalc="1"} is specified in the model XML file,
{\tt ctlike} will also compute the Test Statistics (TS) value that is defined by
\begin{equation}
\mathrm{TS} = 2 \ln L(M) - 2 \ln L(M_{-j}) ,
\label{eq:ts}
\end{equation}
where $L(M)$ is the maximum likelihood value for the full model $M$ and
$L(M_{-j})$ is the maximum likelihood value for a model from which the component $j$ has been
removed.
Under the hypothesis that the model $M$ provides a satisfactory fit of the data, TS
follows a $\chi^2_p$ distribution with $p$ degrees of freedom, where $p$ is the number of 
model parameters in component $j$ \citep{cash1979}.

{\tt ctlike} creates an output model XML file that contains the values of the best fitting model
parameters.
For all free parameters, an {\tt error} attribute is added that provides the statistical uncertainty
in the parameter estimate.
If for a model component the computation of the TS value has been requested, a {\tt ts} attribute
providing the TS value is added.
The output model can be used as an input model for other ctools.

\subsubsection{ctbutterfly}

\begin{figure}
\centering
\includegraphics[width=9cm]{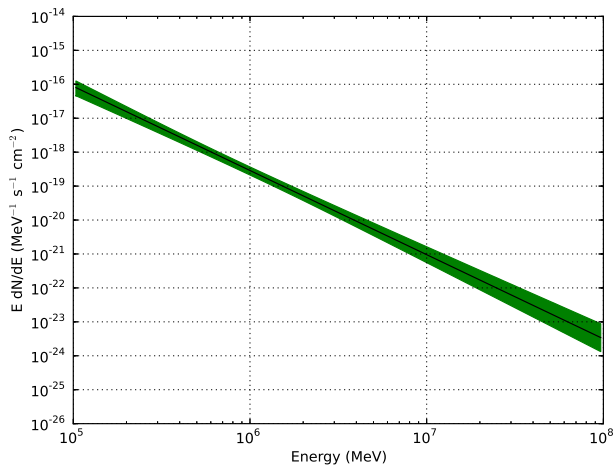}
\caption{
Butterfly diagram for a simulated source with a flux of 10~mCrab, observed with the southern
CTA array for 30 minutes.
\label{fig:butterfly}
}
\end{figure}

The {\tt ctbutterfly} tool calculates a butterfly diagram for a specific source with power law spectral
model.
The butterfly diagram is the envelope of all power law models that are within a given confidence
limit compatible with the data (by default a confidence level of $68\%$ is used).
The tool computes the envelope by evaluating for each energy the minimum and maximum intensity
of all power law models that fall within the error ellipse of the prefactor and index parameters.
The error ellipse is derived from the covariance matrix of a maximum likelihood fit.
The butterfly diagram can be displayed using the {\tt show\_butterfly.py} script that is provided
with the ctools package.
For illustration, Fig.~\ref{fig:butterfly} shows the output of the  {\tt show\_butterfly.py} script, obtained
for a simulated source with a flux of 10~mCrab, observed with the southern CTA array for 30 minutes.

\subsubsection{ctulimit}

The {\tt ctulimit} tool computes the upper flux limit for a specific source model.
Except for the node function, all spectral models are supported.
Starting from the maximum likelihood model parameters, the tool finds the model flux that leads
to a decrease of the likelihood that corresponds to a given confidence level
(by default a confidence level of $95\%$ is used).
{\tt ctulimit} writes the differential upper flux limit at a given reference energy and the integrated
upper flux limit into the log file.

\subsubsection{cterror}

The {\tt cterror} tool computes the parameter errors for a specific source model using the
likelihood profiles.
Starting from the maximum likelihood model parameters, the tool finds the minimum and
maximum model parameters that lead to a decrease of the likelihood that corresponds to
a given confidence level
(by default a confidence level of $68\%$ is used).
{\tt cterror} creates an output model XML file that contains the values of the best fitting model
parameters.
For all free parameters, an {\tt error} attribute is added that provides the statistical uncertainty
in the parameter estimate as obtained from the likelihood profile.
While {\tt cterror} computes asymmetrical errors, which are written into the log file, the XML file
will contain the mean error that is obtained by averaging the negative and positive parameter errors.

\subsubsection{cttsmap}

\begin{figure}
\centering
\includegraphics[width=9cm]{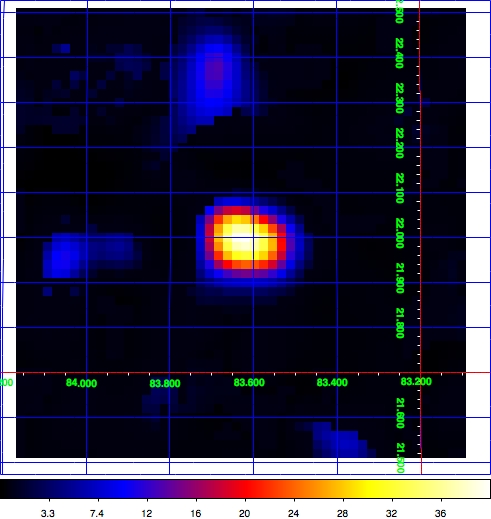}
\caption{
TS map for a simulated source with a flux of 10~mCrab, observed with the southern CTA array
for 30 minutes.
\label{fig:tsmap}
}
\end{figure}

The {\tt cttsmap} tool generates a TS map for a specific source model with point source, radial or
elliptical spatial component.
The tool displaces the specified source on a grid of sky positions and computes for each position
the TS value (see Eq.~\ref{eq:ts}) by fitting the remaining free parameters of the source
model.
If the only remaining free parameter of the source model is the source flux, the
square-root of the TS values corresponds to the pre-trial detection significance of the source in
Gaussian sigma.
We note that the TS values are only valid if at least a few events are actually detected towards a grid
position, which may not always be the case for high energies and/or short observing times
(irrespectively of whether these events actually come from the source or from
instrumental background).
{\tt cttsmap} creates a FITS file comprising a sky map of TS values, followed by one image
extension per free parameter that contain sky maps of the fitted parameter values.
Figure \ref{fig:tsmap} shows a TS map that has been obtained for a simulated source with a flux of
10~mCrab, observed with the southern CTA array for 30 minutes.
Events between $0.1$ and $100$~TeV have been considered.

\subsubsection{ctmodel}

The {\tt ctmodel} tool generates a model cube based on a model definition XML file
(see Appendix \ref{sec:modxml}).
A model cube is a three-dimensional data cube providing the number of predicted counts for a
model as function of reconstructed Right Ascension or Galactic longitude, Declination or Galactic
latitude, and energy. 
{\tt ctmodel} creates a model cube FITS file comprising the predicted number of events per bin,
the energy boundaries of the model cube, and the Good Time Intervals of all observations that
have been used to compute the model cube.

\subsubsection{csresmap}

The {\tt csresmap} scripts generates a residual map for a given model.
It works for event lists, counts cubes or observation definition XML files.
For event lists, parameters that define the spatial and spectral binning need to be provided so
that the script can bin the data internally.
The model is then convolved with the instrumental response function for that binning and
used for residual computation.
Before residual computation, the counts and model cubes are collapsed into maps by summing
over all energies.
Three options exist then for residual computation:
the subtraction of the model from the counts ({\tt algorithm=SUB}),
the subtraction and division by the model ({\tt algorithm=SUBDIV}), and
the subtraction and division by the square root of the model ({\tt algorithm=SUBDIVSQRT}).
By default {\tt algorithm=SUBDIV} is applied.
{\tt csresmap} creates a FITS file containing a sky map of the residuals.

\subsubsection{ctskymap}

\begin{figure}
\centering
\includegraphics[width=9cm]{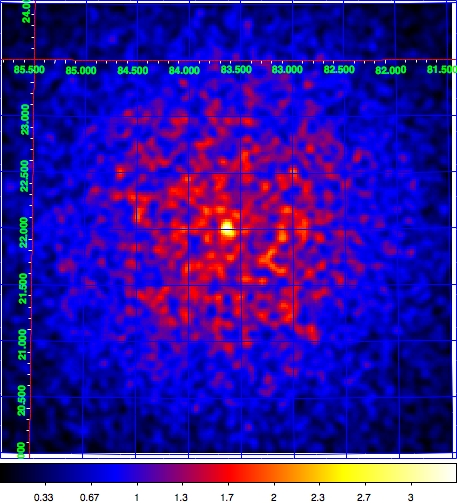}
\caption{
Sky map of the observed events above 100~GeV for a simulated source with a flux of 10~mCrab, 
observed with the southern CTA array for 60 minutes. We note that the map is not background
subtracted.
\label{fig:skymap}
}
\end{figure}

The {\tt ctskymap} tool generates a sky map from either a single event list or the event lists
specified in an observation definition XML file.
The tool will loop over all event lists that are provided and fill all events into a single sky map.
So far, only the generation of maps of the measured number of counts are supported.
{\tt ctskymap} creates a FITS file comprising a sky map.
Figure \ref{fig:skymap} shows a sky map created by {\tt ctskymap} for an observation of a
simulated source with a flux of 10~mCrab, observed with the southern CTA array for 60 minutes.

\subsubsection{csspec}

\begin{figure}
\centering
\includegraphics[width=9cm]{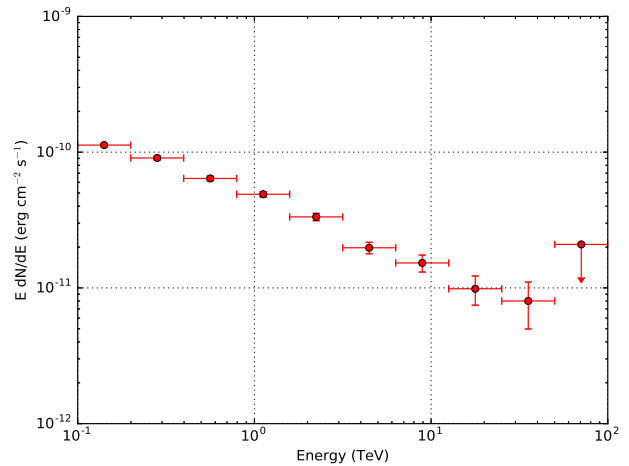}
\caption{
Spectrum obtained using {\tt csspec} for a simulated source with a flux of 1~Crab, observed with
the southern CTA array for 30 minutes.
\label{fig:spectrum}
}
\end{figure}

The {\tt csspec} script extracts the spectrum of a gamma-ray source by fitting a model in a given
set of spectral bins to the data.
The model fits are performed using {\tt ctlike} and the script computes the source flux and its
statistical uncertainty in each spectral bin, as well as the significance of the source detection.
Optionally, {\tt csspec} computes also upper flux limits for each spectral bin.
The script works on event list(s) or counts cube(s).
{\tt csspec} creates a FITS file containing a table with the fitted source spectrum, comprising one
row per spectral bin.
The spectrum can be displayed using the {\tt show\_spectrum.py} script that is provided
with the ctools package.
For illustration, Fig.~\ref{fig:spectrum} shows the output of the script that was obtained
for a simulated source with a flux of 1~Crab, observed with the southern CTA array for 30
minutes.

\subsubsection{cslightcrv}

The {\tt cslightcrv} script computes a light curve by performing a maximum likelihood fit using
{\tt ctlike} in a series of time bins.
The time bins can be either specified in an ASCII file, as an interval divided into equally sized time
bins, or can be taken from the Good Time Intervals of the observation(s).
The format of the ASCII file is one row per time bin, each specifying the start of stop value of the
bin, separated by a whitespace.
Times are specified in Modified Julian Days (MJD).
{\tt cslightcrv} creates a FITS file containing a table with the fitted model parameters and their
statistical errors, the statistical significance of the detection as expressed by the TS value,
and the upper flux limit, with one row per time bin.

\subsubsection{cssens}

\begin{figure}
\centering
\includegraphics[width=9cm]{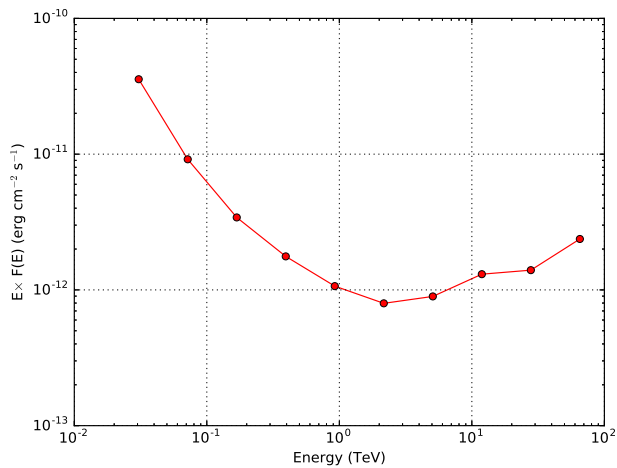}
\caption{
On-axis differential sensitivity for a point source obtained using the {\tt cssens} script for the
southern CTA array (obtained for an observation time of 30 minutes).
\label{fig:sensitivity}
}
\end{figure}

The {\tt cssens} script computes the differential or integrated CTA sensitivity using maximum
likelihood fitting of a test source.
The differential sensitivity is determined for a number of energy bins, the integral sensitivity is
determined for a number of energy thresholds and an assumed source spectrum.
The test source is fitted to simulated data using {\tt ctlike} to determine its detection significance
as a function of source flux.
The source flux is then varied until the source significance achieves a given level, specified by 
the significance parameter {\tt sigma}, and set by default to $5\sigma$.
To reduce the impact of variations between individual Monte Carlo simulations, a sliding average
is applied in the significance computation.
The significance is estimated using the TS value defined by Eq.~(\ref{eq:ts}).
The simplified assumption is made that the significance (in Gaussian sigma) is the square root
of the TS values.
The sensitivity curve can be displayed using the {\tt show\_sensitivity.py} script that is provided
with the ctools package.
Figure \ref{fig:sensitivity} shows the differential sensitivity curve that has been obtained using
the {\tt cssens} script for the southern CTA array after 30 minutes of observations.
Please note that the high-energy sensitivity is slightly better than the one published by
CTA,\footnote{\url{https://portal.cta-observatory.org/Pages/CTA-Performance.aspx}} as the
latter includes an additional constraint of detecting a minimum of 10 photons from a source,
while for a TS computation the presence of a few photons may be sufficient to reach a significance
of $5\sigma$.

\subsubsection{cspull}

The {\tt cspull} script generates pull distributions for all free model parameters.
The pull is defined by
\begin{equation}
g = \frac{x - \mu}{\sigma}
\end{equation}
where
$x$ is the fitted model parameter,
$\mu$ is its true value, and
$\sigma$ is the statistical uncertainty in the fitted model parameter.
For an unbiased and correct estimate of the model parameter and its statistical error, the
pull will be distributed as a standard Gaussian with mean zero and unit width.
The {\tt cspull} script will perform {\tt ntrials} statistically independent computations of the
pull for each free model parameter by simulating events using {\tt ctobssim} followed by a
maximum likelihood model fitting using {\tt ctlike}.
From the output file, pull distribution plots can be generated using the {\tt show\_pull\_histogram.py}
script that is included in the package.
Another script named {\tt show\_pull\_evolution.py} can be used to plot the evolution of the mean
and standard deviation of the pull distribution as function of the number of trials.

\subsubsection{cstsdist}

The {\tt cstsdist} script generates TS distributions (see Eq.~\ref{eq:ts}) for a given source by
repeatedly computing the TS value for {\tt ntrials} simulated data sets.
This script supports unbinned or binned data, support for a stacked analysis is not yet implemented.

\section{Performance}
\label{sec:performance}

\subsection{Numerical accuracy}
\label{sec:accuracy}

The GammaLib and ctools packages have been developed with the goal to achieve an accuracy
of better than $1\%$ in all numerical computations.
This means that if ctools are used to determine for example the flux received from a gamma-ray
source or the spectral points of a spectral energy distribution (SED), the relative precision of the
flux or the spectral points is better than $1\%$.
The same is true for spatial parameters, such as source position or source extension.
For many cases the actual numerical precision is in fact much better than $1\%$, but in any case,
it should never be worse.
Note, however, that this does not imply that source parameters can be determined with an IACT
with an accuracy of $1\%$.
The accuracy depends in the end on the precision to which the instrument response function is 
known, which is currently more in the $10-20\%$ range.

The user should also be aware that bins are always evaluated at their centre, which can lead to
biases when the binning is chosen too coarse.
This is particularly important when performing a binned or stacked analysis, where the spatial and
spectral binning needs to be sufficiently fine grained to fully sample the variation of the model.
In particular, the spatial binning should be better than the best angular resolution over the energy
range of interest.
Typically, a value of $0.02^\circ$ per pixel should be used for the spatial binning, and at least
10 bins per decade for the spectral binning of IACT data.

There is an issue with the fit of the broken power law spectral model, which has unreliable
statistical errors, specifically for the prefactor and the break value.
Errors are in general too large, which is related to the fact that the model gradient is
discontinuous in energy.

The user should also avoid fitting the pivot energies $E_0$ of spectral models.
The pivot energy is not an independent parameter of the spectral models, and consequently, in
case all other spectral parameters are free, the pivot energy is unconstrained.
The pivot energy should therefore always be kept fixed, or other parameters of the spectral
model need to be fixed to assure the non-degeneracy of the free model parameters.

Finally, we note that when the width of the radial shell model becomes comparable to or smaller than
the angular resolution, the shell width tends to be overestimated while the shell radius tends to
be underestimated.
The fitted shell width and radius should thus not be over-interpreted when the width is close to
the angular resolution of the IACT.
Also we note that the convergence of the elliptical Gaussian model can be slow, and in some
situations requires on the order of 20 iterations before the fit converges.
Nevertheless, the numerical accuracy of the model fitting results for the elliptical Gaussian model
is satisfactory.

\subsection{Science verification}

\begin{figure}
\centering
\includegraphics[width=4.4cm]{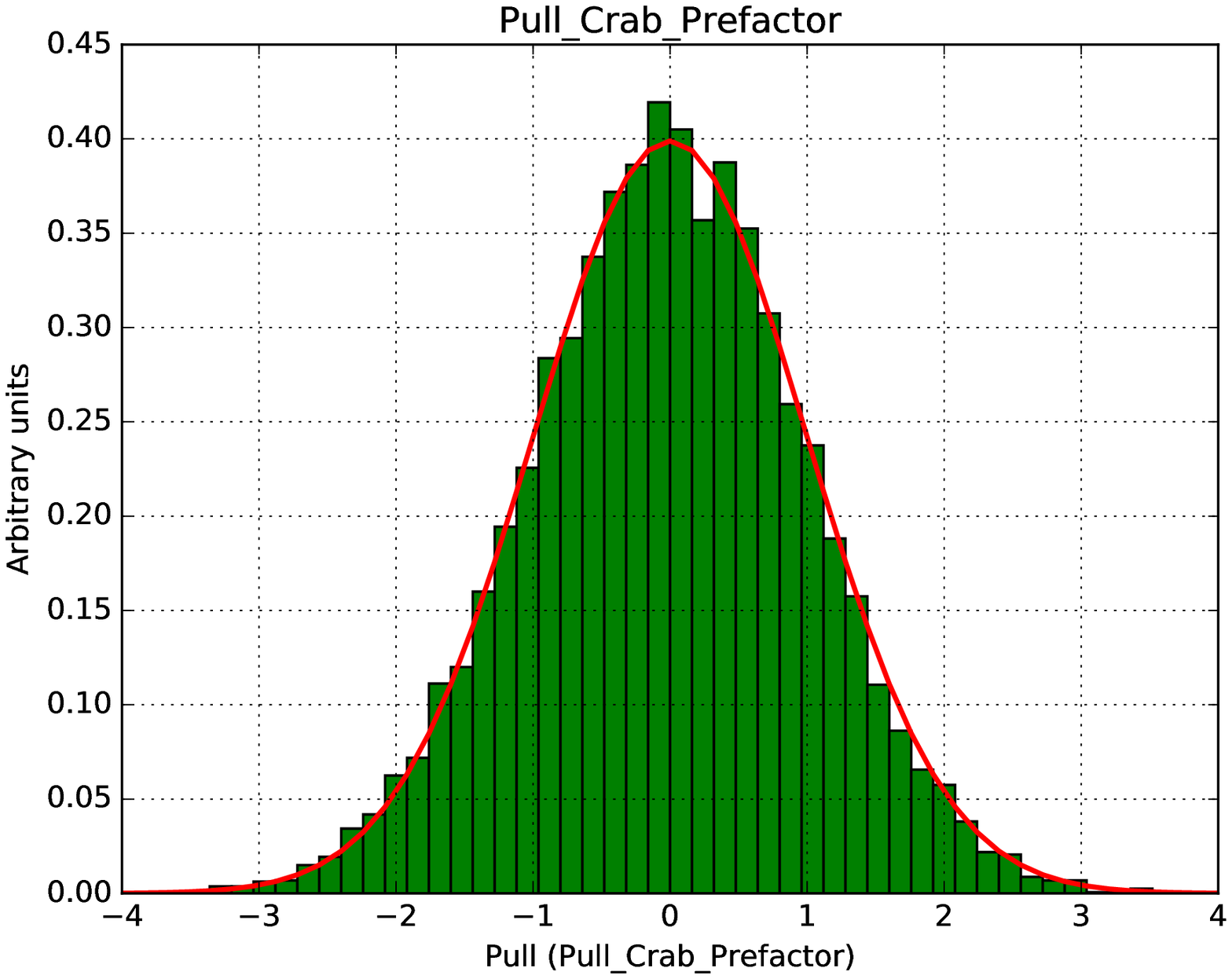}
\includegraphics[width=4.4cm]{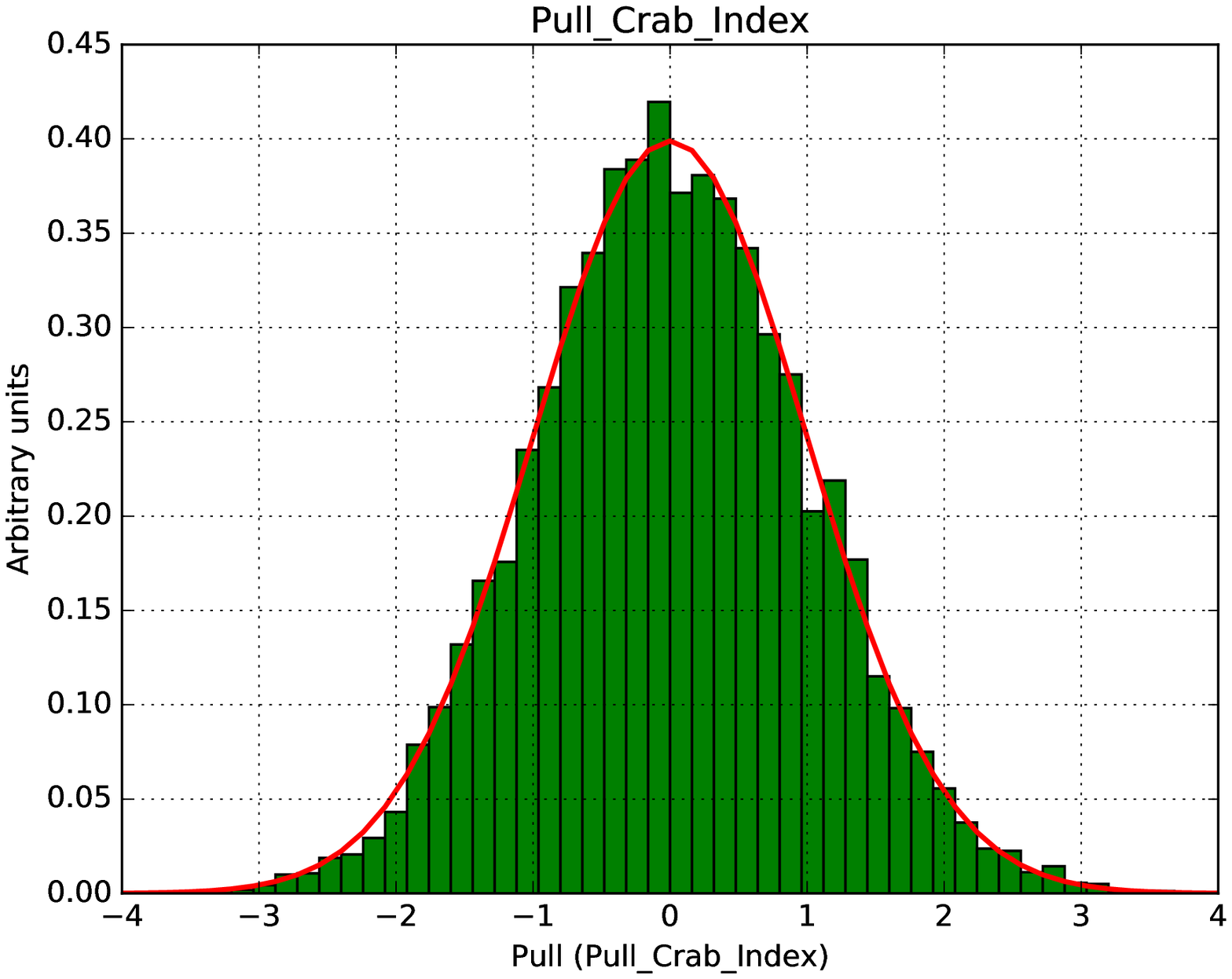}
\caption{
Pull distributions for the prefactor and the index of a power law spectral model for an unbinned
analysis.
\label{fig:pull}
}
\end{figure}

Science verification of GammaLib and ctools is performed by verifying that the pull distributions
of all spatial and spectral models follow a standard Gaussian with mean zero and unit
width.
This verification is done using the {\tt cspull} script.
Figure \ref{fig:pull} shows as an example the pull distribution of the prefactor and the index of
a power law spectral model, obtained for 10\,000 trials and based on a Crab-like point source
observed during an observation duration of 30 minutes for an unbinned analysis.

Science verification is also actively ongoing within collaborations of existing IACTs.
The confrontation of the software to real IACT data is fundamental in verifying the concepts
that have been implemented.
They will also play a key role in validating the handling of instrumental backgrounds using parametric
models, and feed back into their improvements.
Initial results are promising, but since the data is proprietary to the respective collaborations,
we will not publish them here.

\subsection{Benchmarking}

\begin{table}
\caption{Benchmark of ctools unbinned, binned and stacked analysis pipelines as function
of the observation duration.
Quoted numbers give the spent processor time in seconds.
Processor times are quoted for the full pipeline and for the {\tt ctlike} maximum likelihood fitting
step only.}
\label{tab:benchmark}
\centering
\begin{tabular}{c c c c c c c}
\hline\hline
Duration & \multicolumn{2}{c}{Unbinned} & \multicolumn{2}{c}{Binned} & \multicolumn{2}{c}{Stacked} \\
  & full & {\tt ctlike} & full & {\tt ctlike} & full & {\tt ctlike} \\
(h) & (s) & (s) & (s) & (s) & (s) & (s) \\
\hline
 0.5 &     7 &      1 & 41 & 35 & 35 & 28 \\
     1 &    8 &      2 & 41 & 35 & 35 & 27 \\
     2 &    9 &      3 & 42 & 35 & 36 & 28 \\
    4 &   13 &     6 &  43 & 36 & 36 & 27 \\
    8 &   17 &     8 &  44 & 36 & 38 & 28 \\
  16 &   28 &   16 & 48 & 37 & 42 & 29 \\
  32 &   51 &   31 & 53 & 37 & 48 & 29 \\ 
  64 &   94 &    61 & 64 & 37 & 60 & 29 \\
128 & 184 & 124 & 86 & 38 & 86 & 30 \\
256 & 359 & 244 & 129 & 38 & 131 & 31 \\ 
\hline
\end{tabular}
\end{table}

Computing benchmarks of simple ctools analysis pipelines have been generated using the
{\tt benchmark\_cta\_analysis.py} script that is part of the ctools package.
The benchmarks have been obtained on a Dell PowerEdge R815 server equipped with four
12-core AMD Opteron 6174 2.2~GHz processors running a CentOS 5 operating system
(no parallel computing has been used in the pipeline setup).
The unbinned pipeline executes in sequence the {\tt ctobssim}, {\tt ctselect}, and {\tt ctlike} tools
for a region of interest of $3^\circ$.
The binned pipeline executes the {\tt ctobssim}, {\tt ctbin}, and {\tt ctlike} tools for $200\times200$
spatial bins of $0.02^\circ \times 0.02^\circ$ in size, and 40 energy bins.
The stacked pipeline executes the {\tt ctobssim}, {\tt ctbin}, {\tt ctexpcube}, {\tt ctpsfcube},
{\tt ctbkgcube}, and {\tt ctlike} tools for the same binning.
The source model consisted of a single point source with a power law spectrum on top of the
instrumental background.
Only the spectral parameters of the source and the background model have been fitted.
The simulation was done for the southern CTA array using the {\tt South\_50h} IRF.
Events in the energy range 100 GeV -- 100 TeV were simulated and analysed.

\begin{figure}
\centering
\includegraphics[width=9cm]{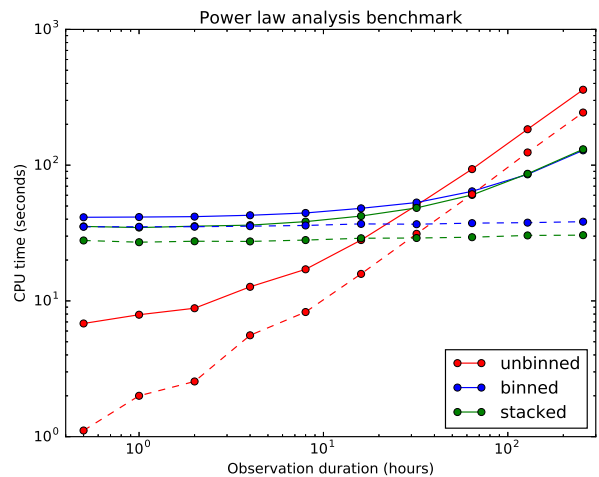}
\caption{
Graphical representation of the ctools computing benchmark results for the fit of the spectrum
of a single point source.
Solid lines give the total time spent in the pipelines while dashed lines represent the time spent
in the {\tt ctlike} tool only.
\label{fig:benchmark}
}
\end{figure}

\begin{table*}
\caption{Benchmark of ctools unbinned, binned and stacked analysis pipelines for an observation
duration of 32 hours as function of the spectral and spatial model.
The first column gives the model type (including the size parameters of the spatial model components),
the second columns indicates the observation duration from which on a stacked analysis is faster 
than an unbinned analysis.
The remaining numbers give the spent processor time in seconds.
Processor times are quoted for the full pipeline and for the {\tt ctlike} maximum likelihood fitting step
only.
}
\label{tab:benchmark_model}
\centering
\begin{tabular}{l c c c c c c c}
\hline\hline
Model & Unbinned/stacked & \multicolumn{2}{c}{Unbinned} & \multicolumn{2}{c}{Binned} & \multicolumn{2}{c}{Stacked} \\
 & crossover & full & {\tt ctlike} & full & {\tt ctlike} & full & {\tt ctlike} \\
 & duration (h) & (s) & (s) & (s) & (s) & (s) & (s) \\
\hline
Power law & 30 &   51 &  31 & 53 & 37 & 48 & 29 \\
Power law 2 & 25 & 53 & 33 & 57 & 40 & 49 & 29 \\
Broken power law & 7 & 201 & 183 & 76 & 60 & 54 & 35 \\
Exponentially cut-off power law & 22 & 65 & 40 & 60 & 39 & 54 & 30 \\
Super exponentially cut-off power law & 6 & 242 & 166 & 117 & 51 & 102 & 31 \\
Log parabola & 7 & 204 & 125 & 106 & 38 & 108 & 34 \\
File function & 30 & 50 & 31 & 55 & 39 & 51 & 31 \\
Node function & 33 & 48 & 30 & 57 & 41 & 53 & 35 \\
\hline
Point source & 30 & 97 & 78 & 127 & 111 & 89 & 70 \\
Isotropic diffuse source & 22 & 590 & 478 & 3515 & 3407 & 469 & 358 \\
Sky map source ($4^\circ \times 9^\circ$ map) & 0.5 & 613 & 595 & 4434 & 4419 & 43 & 25 \\
Map cube source ($1^\circ \times 1^\circ$ map) & 22 & 922 & 900 & 5053 & 5043 & 662 & 640 \\
Radial disk source ($r=0.2^\circ$) & 2 & 3289 & 3268 & 592 & 574 & 320 & 299 \\
Radial Gaussian source ($\sigma=0.2^\circ$) & 15 & 19089 & 19068 & 3404 & 3386 & 2683 & 2661 \\
Radial shell source ($\theta_\mathrm{in}=0.3^\circ$, $\theta_\mathrm{out}=0.4^\circ$) & 
  3 & 32530 & 32509 & 8439 & 8420 & 1327 & 1305 \\
Elliptical disk source ($a=0.2^\circ$, $b=0.4^\circ$) &
  6 & 11864 & 11840 & 2324 & 2302 & 2893 & 2869 \\
Elliptical Gaussian source ($a=0.2^\circ$, $b=0.4^\circ$) &
  3 & 154337 & 154316 & 6559 & 6540 & 9904 & 9882 \\
\hline
\end{tabular}
\end{table*}

Table \ref{tab:benchmark} summarises the spent processor time in seconds in each of the analysis
pipelines for observation durations ranging from 30 minutes to 256 hours.
Figure \ref{fig:benchmark} provides a graphical representation of the results, where solid lines
give the total time spent in the pipelines while dashed lines represent the time spent in the
{\tt ctlike} tool only.
The processor time for {\tt ctlike} in a binned or stacked analysis is essentially independent of the
duration of the observation, which is expected since the number of operations for these analyses
depend only on the number of bins and not the number of events (see Eq.~\ref{eq:binned}).
In contrast, for an unbinned analysis the number of required operations increases with the
number of events (see Eq.~\ref{eq:unbinned}), and the processor time increases accordingly.
The crossover where the time spent in {\tt ctlike} is equal for unbinned and binned or stacked
analysis is for an observation duration of $\sim30$ hours.
The additional processor time for the full pipeline comes essentially from the event simulation
step using {\tt ctobssim}, which in all cases increases with the duration of the observing time.
Above $\sim60$ hours, the time spent in event simulation exceeds the time spent in maximum
likelihood model fitting for a binned or stacked analysis.
For unbinned analysis, the time spent in maximum likelihood fitting dominates for observation
durations longer than $\sim5$ hours.

Processor time benchmarks have also been obtained for all spectral and spatial models.
Table \ref{tab:benchmark_model} summarises the results for an observation duration of
32 hours.
The first part of the table provides benchmarks for various spectral models that have been
combined with a point source spatial component with fixed position.
While the performance of the stacked analysis is relatively insensitive of the spectral model
that is used, unbinned and binned analyses show some variations.
The broken power law, super exponentially cut off power law, and the log parabola models
take longest in an unbinned analysis, making a stacked analysis more efficient for observation
durations longer than $6-7$ hours.
For the broken power law, an unbinned analysis takes substantially more fit iterations than a
binned or a stacked analysis, which is related to the discontinuity of the model gradients
mentioned above (cf.~section \ref{sec:accuracy}). 
The super exponentially cut-off power law and the log parabola model are expensive in the
evaluation, which is a drawback for unbinned analysis where the model is evaluated for each
event, in contrast to a binned or stacked analysis, where the model is effectively only evaluated
once for each energy layer thanks to an internal value caching mechanism.

The second part of table \ref{tab:benchmark_model} summarises benchmarks for various
spatial models that have been combined with a power law spectral model with free prefactor
and index.
All spatial parameters (position, size or orientation) have been adjusted by the fit.
Fits of diffuse models take generally somewhat longer than fits of a point source, the only
exception being the use of a sky map in a stacked analysis which is relatively fast due to an
efficient internal caching of response values.
We note that all diffuse model response computations benefit from an internal caching mechanism
that evaluates the response only once for each event or event bin, taking advantage of the absence
of spatial parameters that modify the emission morphology for diffuse models.
Fits of radial or elliptical models take generally substantially longer than fits of a point source
or of diffuse models, owing to the larger number of parameters that need to be adjusted.
As spatial parameters have no analytical gradient, the derivatives have to be computed
numerically, multiplying by three the number of function evaluations that are needed for each
spatial parameter.
Also, no simple response caching can be performed.
The large number of iterations needed for the convergence of the elliptical Gaussian
source (see section \ref{sec:accuracy}) explains why the processing time is excessively large
for this spatial model.
Nevertheless, future code developments will aim at reducing the processing times for the radial 
and elliptical spatial models, as they will be central to many science analyses of IACT data. 

For the time being, a stacked analysis is faster than an unbinned analysis for observation durations 
that are longer than a few hours, although the precise crossover duration is very model dependent.
With the exception of the elliptical Gaussian source model, the {\tt ctlike} processing time for
a stacked analysis is always less than an hour, and can be as fast as 5 minutes for a radial disk 
model.
A stacked analysis is therefore very often the method of choice for fitting radial or elliptical spatial
models to IACT data.

\section{Outlook}
\label{sec:outlook}

Following several years of development and testing, we have now released the GammaLib
and ctools software packages for the scientific analysis of gamma-ray event data.
So far, the software mainly targets the analysis of Imaging Air Cherenkov Telescope event
data, but it can also be used for the analysis of Fermi-LAT data or the exploitation of the
COMPTEL legacy archive.

Future developments will focus on the expansion of the ctools package to cover all features 
that are needed for a fully developed CTA Science Tools software package.
This includes support for classical very-high-energy analysis techniques,
such as aperture photometry and on-off fitting,
the addition of tools for source extraction and identification, as well as for the timing
analysis of pulsars and binaries.
It is also planned to enhance the support for imaging analysis, for example by adding a tool
that enables the spatial deconvolution of the data to produce super-resolved sky maps.

The GammaLib package will be expanded to also cover the analysis of Pass 8 data for Fermi-LAT,
to enhance the COMPTEL module by an improved treatment of the instrumental background,
and to implement full interoperability with Virtual Observatory tools.
Including additional instrument modules to broaden the telescope coverage is also under
investigation.

We want to conclude with the reminder that ctools and GammaLib are open source software
packages that benefit from a dynamic and enthusiastic community of researchers.
The projects are open to new contributions and enhancements, and if you have suggestions
or ideas on how to expand the existing software, you are warmly welcomed to join the
development team.

\begin{acknowledgements}
We would like to acknowledge highly valuable discussions with members of
the CTA Consortium,
the Fermi-LAT Collaboration,
the H.E.S.S. Collaboration,
the VERITAS Collaboration, and
the MAGIC Collaboration
that all contributed to improve and consolidate the GammaLib and ctools software packages.
We also want to thank W. Collmar for making the COMPTEL instrument response functions 
available for inclusion into GammaLib.
This work has been carried out thanks to the support of the OCEVU Labex (ANR-11-LABX-0060)
and the A$^\ast$MIDEX project (ANR-11-IDEX-0001-02) funded by the ``Investissements d'Avenir''
French government program managed by the ANR.
\end{acknowledgements}

\noindent
This paper has gone through internal review by the CTA Consortium.

%
%

\bibliographystyle{aa} 
\bibliography{references} 

\begin{thebibliography}{17}
\expandafter\ifx\csname natexlab\endcsname\relax\def\natexlab#1{#1}\fi

\bibitem[{{Acharya} {et~al.}(2013){Acharya}, {Actis}, {Aghajani}, {Agnetta},
  {Aguilar}, {Aharonian}, {Ajello}, {Akhperjanian}, {Alcubierre},
  {Aleksi{\'c}}, \& et~al.}]{acharya2013}
{Acharya}, B.~S., {Actis}, M., {Aghajani}, T., {et~al.} 2013, Astroparticle
  Physics, 43, 3

\bibitem[{{Arnaud}(1996)}]{arnaud1996}
{Arnaud}, K.~A. 1996, in Astronomical Society of the Pacific Conference Series,
  Vol. 101, Astronomical Data Analysis Software and Systems V, ed. G.~H.
  {Jacoby} \& J.~{Barnes}, 17

\bibitem[{{Cash}(1979)}]{cash1979}
{Cash}, W. 1979, \apj, 228, 939

\bibitem[{{Corcoran} {et~al.}(1995){Corcoran}, {Angelini}, {George}, {McGlynn},
  {Mukai}, {Pence}, \& {Rots}}]{corcoran1995}
{Corcoran}, M.~F., {Angelini}, L., {George}, I., {et~al.} 1995, in Astronomical
  Society of the Pacific Conference Series, Vol.~77, Astronomical Data Analysis
  Software and Systems IV, ed. {R.~A.~Shaw, H.~E.~Payne, \& J.~J.~E.~Hayes},
  219--+

\bibitem[{{de Boer} {et~al.}(1992){de Boer}, {Bennett}, {Bloemen}, {den
  Herder}, {Hermsen}, {Klumper}, {Lichti}, {McConnell}, {Ryan},
  {Sch{\"o}nfelder}, {Strong}, \& {de Vries}}]{deboer1992}
{de Boer}, H., {Bennett}, K., {Bloemen}, H., {et~al.} 1992, in Data Analysis in
  Astronomy, ed. V.~{di Gesu}, L.~{Scarsi}, R.~{Buccheri}, \& P.~{Crane},
  241--249

\bibitem[{{Diehl} {et~al.}(2003){Diehl}, {Baby}, {Beckmann}, {Connell},
  {Dubath}, {Jean}, {Kn{\"o}dlseder}, {Roques}, {Schanne}, {Shrader},
  {Skinner}, {Strong}, {Sturner}, {Teegarden}, {von Kienlin}, \&
  {Weidenspointner}}]{diehl2003}
{Diehl}, R., {Baby}, N., {Beckmann}, V., {et~al.} 2003, \aap, 411, L117

\bibitem[{{G{\'o}rski} {et~al.}(2005){G{\'o}rski}, {Hivon}, {Banday},
  {Wandelt}, {Hansen}, {Reinecke}, \& {Bartelmann}}]{gorski2005}
{G{\'o}rski}, K.~M., {Hivon}, E., {Banday}, A.~J., {et~al.} 2005, \apj, 622,
  759

\bibitem[{{Greisen} \& {Calabretta}(2002)}]{greisen2002}
{Greisen}, E.~W. \& {Calabretta}, M.~R. 2002, \aap, 395, 1061

\bibitem[{Hunter(2007)}]{hunter2007}
Hunter, J.~D. 2007, Computing In Science \& Engineering, 9, 90

\bibitem[{{Kn{\"o}dlseder} {et~al.}(2015){Kn{\"o}dlseder}, {Beckmann},
  {Boisson}, {Brau-Nogu{\'e}}, {Deil}, {Kh{\'e}lifi}, {Mayer}, {Walter}, \&
  {CTA Consortium}}]{knoedlseder2015}
{Kn{\"o}dlseder}, J., {Beckmann}, V., {Boisson}, C., {et~al.} 2015, ArXiv
  e-prints [\eprint[arXiv]{1508.06078}]

\bibitem[{{Kn{\"o}dlseder} {et~al.}(2016){Kn{\"o}dlseder}, {Mayer}, {Deil},
  {Buehler}, {Bregeon}, \& {Martin}}]{knoedlseder2016}
{Kn{\"o}dlseder}, J., {Mayer}, M., {Deil}, C., {et~al.} 2016, Astrophysics
  Source Code Library [\eprint{ascl:1601.005}]

\bibitem[{{Kn{\"o}dlseder} {et~al.}(2011){Kn{\"o}dlseder}, {Mayer}, {Deil},
  {Cayrou}, {Owen}, {Kelley-Hoskins}, {Lu}, {Buehler}, {Forest}, {Louge},
  {Siejkowski}, {Kosack}, {Gerard}, {Schulz}, {Martin}, {Hassan}, \&
  {Brau-Nogu'e}}]{knoedlseder2011}
{Kn{\"o}dlseder}, J., {Mayer}, M., {Deil}, C., {et~al.} 2011, Astrophysics
  Source Code Library [\eprint{ascl:1110.007}]

\bibitem[{{Marquardt}(1963)}]{marquardt1963}
{Marquardt}, D.~W. 1963, Journal of the Society for Industrial and Applied
  Mathematics, 11, 431

\bibitem[{{Marsaglia} \& {Zaman}(1994)}]{marsaglia1994}
{Marsaglia}, G. \& {Zaman}, A. 1994, Computers in Physics, 8, 117

\bibitem[{{Mattox} {et~al.}(1996){Mattox}, {Bertsch}, {Chiang}, {Dingus},
  {Digel}, {Esposito}, {Fierro}, {Hartman}, {Hunter}, {Kanbach}, {Kniffen},
  {Lin}, {Macomb}, {Mayer-Hasselwander}, {Michelson}, {von Montigny},
  {Mukherjee}, {Nolan}, {Ramanamurthy}, {Schneid}, {Sreekumar}, {Thompson}, \&
  {Willis}}]{mattox1996}
{Mattox}, J.~R., {Bertsch}, D.~L., {Chiang}, J., {et~al.} 1996, \apj, 461, 396

\bibitem[{{Pence} {et~al.}(1993){Pence}, {Blackburn}, \& {Greene}}]{pence1993}
{Pence}, W., {Blackburn}, J.~K., \& {Greene}, E. 1993, in Astronomical Society
  of the Pacific Conference Series, Vol.~52, Astronomical Data Analysis
  Software and Systems II, ed. R.~J. {Hanisch}, R.~J.~V. {Brissenden}, \&
  J.~{Barnes}, 541

\bibitem[{{Pence} {et~al.}(2010){Pence}, {Chiappetti}, {Page}, {Shaw}, \&
  {Stobie}}]{pence2010}
{Pence}, W.~D., {Chiappetti}, L., {Page}, C.~G., {Shaw}, R.~A., \& {Stobie}, E.
  2010, \aap, 524, A42+

\end{thebibliography}

\begin{appendix}

\section{Installing the software}

GammaLib is distributed as source code releases that can be downloaded from
\url{http://cta.irap.omp.eu/gammalib/download.html}.
Alternatively, the trunk of the source code can be accessed by cloning from the Git repository
using

{\tiny
\begin{verbatim}
$ export GIT_SSL_NO_VERIFY=true
$ git clone https://cta-gitlab.irap.omp.eu/gammalib/gammalib.git
\end{verbatim}
}

\noindent Building and installing the code follows the standard procedure for GNU software
packages:

{\tiny
\begin{verbatim}
$ ./autogen.sh
$ ./configure
$ make
$ make check
$ [sudo] make install
\end{verbatim}
}

\noindent The first command is only required when the source code has been cloned using
Git, the (optional) {\tt sudo} command is necessary when admin privileges are needed to install
the library into the default {\tt /usr/local/gamma} directory.

All instrument-specific modules are compiled by default when building the package, but can
optionally be disabled during the configuration step.
For instance

{\tiny
\begin{verbatim}
$ ./configure --without-lat --without-com --without-mwl
\end{verbatim}
}

\noindent configures GammaLib to exclude the Fermi-LAT, COMPTEL and multi-wavelength
modules.

The ctools package is distributed as source code releases that can be downloaded from
\url{http://cta.irap.omp.eu/ctools/download.html}.
Alternatively, the trunk of the source code can be accessed by cloning from the Git repository

{\tiny
\begin{verbatim}
$ export GIT_SSL_NO_VERIFY=true
$ git clone https://cta-gitlab.irap.omp.eu/ctools/ctools.git
\end{verbatim}
}

\noindent Building and installing the code follows the same procedure as for GammaLib.
To get started with GammaLib and ctools, refer to the {\em ``Quickstart"} and {\em ``Using ctools from
Python"} sections of the user documentation at \url{http://cta.irap.omp.eu/ctools}.

Science verification is part of the continuous testing strategy implemented for ctools, and a user
can exercise the standard science verification pipeline by typing the

{\tiny
\begin{verbatim}
$ make science-verification
\end{verbatim}
}

\noindent command in the ctools source package after having installed and configured
the software.
Be aware, however, that this standard test will take more than a day for completion.

\section{Observation definition XML format}
\label{sec:obsxml}

The possibility of analysing data from different instruments with the same software framework
opens up the opportunity to use GammaLib and ctools for multi-instrument and multi-wavelength
analyses.
To facilitate multi-instrument event data analysis, the {\tt GObservations} class can directly load
observations based on information provided in an observation definition XML file.
In Python, the syntax for loading a observation definition XML file is

{\tiny
\begin{verbatim}
$ python
>>> import gammalib
>>> obs = gammalib.GObservations("myobservations.xml")
\end{verbatim}
}

\noindent The format of the observation definition XML file is illustrated below:

{\tiny
\begin{verbatim}
<?xml version="1.0" standalone="no"?>
<observation_list title="observation library">
  <observation name="Crab" id="1" instrument="COM">
    <parameter name="DRE" file="m50439_dre.fits"/>
    <parameter name="DRB" file="m34997_drg.fits"/>
    <parameter name="DRG" file="m34997_drg.fits"/>
    <parameter name="DRX" file="m32171_drx.fits"/>
    <parameter name="IAQ" file="ENERG(1.0-3.0)MeV"/>
  </observation>
  <observation name="Crab" id="1" instrument="LAT">
    <parameter name="CountsMap"    file="srcmap.fits"/>
    <parameter name="ExposureMap"  file="expmap.fits"/>
    <parameter name="LiveTimeCube" file="ltcube.fits"/>
    <parameter name="IRF"          value="P7SOURCE_V6"/>
  </observation>
  <observation name="Crab" id="1" instrument="CTA">
    <parameter name="EventList"   file="cta_events.fits"/>
    <parameter name="Calibration" database="prod2"
                                  response="South_0.5h"/>
  </observation>
</observation_list>
\end{verbatim}
}

\noindent The definition of the observations is enclosed in a single {\tt <observation\_list>} element
that can contain an arbitrary number of {\tt <observation>} elements.
In the example above, three {\tt <observation>} elements are present, and the {\tt instrument}
attribute specifies the instrument code for each of these observations.
Based on that attribute, GammaLib appends instrument-specific observations to the
{\tt GObservations} container, and dispatches the reading of the {\tt <observation>} XML elements
to the respective {\tt GObservation::read()} methods that interpret the instrument specific content.

Optionally, energy thresholds for a given CTA observation can be specified using the {\tt emin} and
{\tt emax} attributes.

{\tiny
\begin{verbatim}
<observation name="Crab" id="1" instrument="CTA" 
                                       emin="0.1" emax="100">
\end{verbatim}
}

\noindent The units of the energy threshold are TeV.
The {\tt ctselect} tool will automatically apply these energy thresholds to the data if the parameter
{\tt usethres=USER} is specified.

\section{Model definition XML format}
\label{sec:modxml}

Models can be defined by manipulating the C++ or Python model classes, but alternatively
a model definition can be specified in form of an XML file that can directly be loaded by
the {\tt GModels} class.
In Python, the syntax for loading a model definition XML file is

{\tiny
\begin{verbatim}
$ python
>>> import gammalib
>>> models = gammalib.GModels("mymodels.xml")
\end{verbatim}
}

\noindent The format of the model definition XML file is inspired from, and is compatible with, the 
format used by the Fermi-LAT Science Tools.\footnote{\url{http://fermi.gsfc.nasa.gov/ssc/}}
The general structure of a model definition XML file is

{\tiny
\begin{verbatim}
<?xml version="1.0" standalone="no"?>
<source_library title="source library">
  <source name="Crab" type="PointSource">
    <spectrum type="PowerLaw">
       <parameter name="Prefactor" .../>
       <parameter name="Index"     .../>
       <parameter name="Scale"     .../>
    </spectrum>
    <spatialModel type="SkyDirFunction">
      <parameter name="RA"  .../>
      <parameter name="DEC" .../>
    </spatialModel>
  </source>
  <source name="Bkg" type="CTAIrfBackground" instrument="CTA">
    <spectrum type="PowerLaw">
      <parameter name="Prefactor" .../>
      <parameter name="Index"     .../>
      <parameter name="Scale"     .../>
    </spectrum>
  </source>
</source_library>
\end{verbatim}
}

\noindent The collection of models is contained in a single {\tt <source\_library>} element that
in turn can contain an arbitrary number of {\tt <source>} elements.
In the example above, two {\tt <source>} elements are present which each correspond to one
model.
The {\tt type} attribute of each {\tt <source>} element allows GammaLib to identify the appropriate
model class.
In the example, the first model is a celestial point source model with a power law spectral
component, and the second model is an instrumental background model for CTA.
Each model component contains a number of model parameters of the form

{\tiny
\begin{verbatim}
<parameter name=".." scale=".." value=".." min=".." max=".."
                                                 free=".."/>
\end{verbatim}
}

\noindent where
\begin{itemize}
\item {\tt name} is the model parameter name that is unique within a model component
\item {\tt scale} is a scale factor that will be multiplied with {\tt value} to provide a model
parameter (this pre-scaling allows all {\tt value} attributes to be of the same order,
which is needed to guarantee the numerical stability when the model is fitted to the
data)
\item {\tt value} is the pre-scaled model value (see above)
\item {\tt min} is the lower limit for the {\tt value} term
\item {\tt max} is the upper limit for the {\tt value} term
\item {\tt free} is a flag that specifies whether a model parameter should be fitted ({\tt free="1"})
or kept fixed ({\tt free="0"}) when then model is fitted to the data
\end{itemize}
Please refer to \url{http://cta.irap.omp.eu/gammalib/user_manual/modules/model.html}
for a detailed description of the syntax of the model definition XML file for each model
type.

\section{IRAF command-line parameter interface}
\label{sec:iraf}

GammaLib implements the IRAF command line parameter interface, a format for specifying
application parameters that is already widely used for high-energy astronomy analysis frameworks,
including ftools, the Chandra CIAO package, the INTEGRAL OSA software, or the Fermi-LAT
Science Tools.
Following the IRAF standard, user parameters are specified in a so-called parameter file, which
is an ASCII file that lists one parameter per row in the format

{\tiny
\begin{verbatim}
name, type, mode, value, minimum, maximum, prompt
\end{verbatim}
}

\noindent where
\begin{itemize}
\item {\tt name} specifies a unique user parameter name
\item {\tt type} specifies the type of the user parameter and is one of "b", "i", "r", "s", "f", "fr", "fw", "fe", "fn",
which stands for boolean, integer, real (or floating point), string, and file name. The various file name
types test for read access, write access, file existence, and file absence, respectively.
\item {\tt mode} specifies the user parameter mode and is one of "a", "h", "l", "q", "hl", "ql", "lh", "lq",
where "a" stands for "automatic", "h" for "hidden", "l" for "learn", and "q" for "query" (the other possibilities
are combinations of modes); "automatic" means that the mode is inferred from the "mode"
parameter in the parameter file, "hidden" means that the parameter is not queried, but can be
specified by explicitly setting its value on the command line or from within Python, "l" means that
the value entered by the user will be written to the parameter file on disk, so that the next time it
will be used as the default value, "q" means that the parameter will be queried.
\item {\tt value} specifies the user parameter value. Possible values for boolean parameters are
case-insensitive "y", "n", "yes", "no", "t", "f", "true", and "false". The special case-insensitive
values "indef", "none", "undef" or "undefined" are used to signal that a value is not
defined. If "inf", "infinity" or "nan" is specified and the parameter is of integer type, its value
will be set to the largest possible integer value. If the parameter is of floating point type,
its status will be set to "Not A Number".
\item {\tt minimum} specifies either a list of options, or the minimum user parameter value for integer or floating
point types, provided that {\tt maximum} is also specified. Options are separated by "|" characters.
\item {\tt maximum} specifies for integer or floating point types the maximum user parameter value, provided
that {\tt minimum} is also specified.
\item {\tt prompt} specifies the text that will be prompted when querying for user parameters.
\end{itemize}

\noindent User parameters can be either queried interactively or can be specified on the
command line which inhibits an application to query the specified parameters.
Command line parameters are given by specifying the parameter name, followed without whitespace
by an equality symbol and the parameter value (hyphens are not needed).
For example

{\tiny
\begin{verbatim}
$ ctobssim infile=myfilefits ra=83.0 dec=22.0
\end{verbatim}
}

\noindent will set the {\tt infile}, {\tt ra}, and {\tt dec} parameters of the {\tt ctobssim} tool.
Using

{\tiny
\begin{verbatim}
$ ctobssim debug=yes
\end{verbatim}
}

\noindent will instruct the {\tt ctobssim} to log any output into the console.
And help text about {\tt ctobssim} can be displayed using

{\tiny
\begin{verbatim}
$ ctobssim --help
\end{verbatim}
}

\end{appendix}

\end{document}